\RequirePackage{snapshot}
\RequirePackage{cmap}

\RequirePackage[svgnames]{xcolor}

\documentclass[final]{aa}

\let\oldpageref\pageref
\renewcommand{\pageref}{\oldpageref*}

\usepackage{silence}
\usepackage[varg]{txfonts}
\usepackage[scaled=0.76]{beramono}

\usepackage{amsmath}
\usepackage{amssymb}
\usepackage{bm}
\usepackage{derivative}

\usepackage{natbib}
\bibpunct{(}{)}{;}{a}{}{,}

\let\oldbibliography\thebibliography
\renewcommand{\thebibliography}[1]{%
  \oldbibliography{#1}%
  \setlength{\itemsep}{2.5pt}%
  \setlength{\baselineskip}{7.5pt}
  \setlength{\lineskiplimit}{-\maxdimen}
}

\usepackage{array}
\usepackage{booktabs}
\usepackage{chemformula}
\usepackage{datetime2}
\usepackage[inline]{enumitem}
\usepackage{fontawesome5}
\usepackage{graphicx}
\usepackage{microtype}
\usepackage{multirow}
\usepackage{placeonpage}  %
\usepackage{ragged2e}
\usepackage{standalone}
\usepackage{subcaption}
\usepackage{siunitx}
\usepackage{tabularx}
\usepackage[flushleft]{threeparttable}
\usepackage[hang,flushmargin]{footmisc}

\DeclareSIUnit\epoch{epoch}

\usepackage{tikz}
\usetikzlibrary{
    calc, 
    positioning,
    shapes.geometric,
}
\usepackage{pgfplots}
\pgfplotsset{compat=1.18}

\usepackage[
    colorlinks=true,
    allcolors=Navy,
]{hyperref}
\usepackage{cleveref}

\newcommand{\given}{\ensuremath{\,|\,}}

\newcommand{\extlink}{\kern0.25em\raisebox{0.5pt}{\scalebox{0.6}{\faIcon{external-link-alt}}}}
\newcommand{\codelink}[1]{\href{#1}{\scalebox{0.75}{\faIcon[regular]{file-code}}}}

\newcommand{\petitRADTRANS}{\texttt{petitRADTRANS}\xspace}
\newcommand{\nautilus}{\texttt{nautilus}\xspace}
\newcommand{\MultiNest}{\texttt{MultiNest}\xspace}
\newcommand{\dynesty}{\texttt{dynesty}\xspace}
\newcommand{\UltraNest}{\texttt{UltraNest}\xspace}

\definecolor{CBF0}{HTML}{5790fc}
\definecolor{CBF1}{HTML}{f89c20}
\definecolor{CBF2}{HTML}{e42536}

\usepackage{csquotes}

\begin{document}

    \title{
        Flow Matching for Atmospheric Retrieval of Exoplanets:\\ 
        Where Reliability meets Adaptive Noise Levels
    }
    \author{
        Timothy D. Gebhard\inst{1,2}\fnmsep%
        \thanks{Correspondence: \href{mailto:tgebhard@tue.mpg.de}{tgebhard@tue.mpg.de}.} \and %
        Jonas Wildberger\inst{1,3} \and %
        Maximilian Dax\inst{1} \and %
        Annalena Kofler\inst{1,4} \and\\ %
        Daniel Angerhausen\inst{2} \and %
        Sascha P. Quanz\inst{2,5} \and %
        Bernhard Schölkopf\inst{1,6} %
    }
    \authorrunning{
        Gebhard et al. (2024)
    }
    \titlerunning{
        Flow Matching for Atmospheric Retrieval of Exoplanets
    }
    \institute{
        \inst{1}Max Planck Institute for Intelligent Systems, Max-Planck-Ring 4, 72076 Tübingen, Germany\\
        \inst{2}ETH Zurich, Institute for Particle Physics \& Astrophysics, Wolfgang-Pauli-Strasse 27, 8093 Zurich, Switzerland\\
        \inst{3}ELLIS Institute Tübingen, Maria-von-Linden-Straße 2, 72076 Tübingen, Germany\\
        \inst{4}Max Planck Institute for Gravitational Physics (Albert Einstein Institute), Am Mühlenberg 1, 14476 Potsdam, Germany\\
        \inst{5}ETH Zurich, Department of Earth and Planetary Sciences, Sonneggstrasse 5, 8092 Zurich, Switzerland\\
        \inst{6}ETH Zurich, Department of Computer Science, Universitätsstrasse 6, 8092 Zurich, Switzerland
    }
    \date{Received 12 August 2024 / Accepted 16 October 2024}
    \abstract{
        Inferring atmospheric properties of exoplanets from observed spectra is key to understanding their formation, evolution, and habitability.
        Since traditional Bayesian approaches to atmospheric retrieval (e.g., nested sampling) are computationally expensive, a growing number of machine learning (ML) methods such as neural posterior estimation (NPE) have been proposed.
    }{
        We seek to make ML-based atmospheric retrieval (1)~more reliable and accurate with verified results, and (2)~more flexible with respect to the underlying neural networks and the choice of the assumed noise models.
    }{
        First, we adopt flow matching posterior estimation (FMPE) as a new ML approach to atmospheric retrieval. 
        FMPE maintains many advantages of NPE, but provides greater architectural flexibility and scalability.
        Second, we use importance sampling~(IS) to verify and correct ML results, and to compute an estimate of the Bayesian evidence.
        Third, we condition our ML models on the assumed noise level of a spectrum (i.e., error bars), thus making them adaptable to different noise models.
    }{
        Both our noise level-conditional FMPE and NPE models perform on par with nested sampling across a range of noise levels when tested on simulated data.
        FMPE trains about 3 times faster than NPE and yields higher IS efficiencies.
        IS successfully corrects inaccurate ML results, identifies model failures via low efficiencies, and provides accurate estimates of the Bayesian evidence.
    }{
        FMPE is a powerful alternative to NPE for fast, amortized, and parallelizable atmospheric retrieval.
        IS can verify results, thus helping to build confidence in ML-based approaches, while also facilitating model comparison via the evidence ratio.
        Noise level conditioning allows design studies for future instruments to be scaled up, for example, in terms of the range of signal-to-noise ratios.
    }
    \maketitle

    \begin{figure*}[t]
        \placeonpage{3}
        \centering
        \begin{subcaptionblock}{8.25cm}
            \centering
            \includegraphics{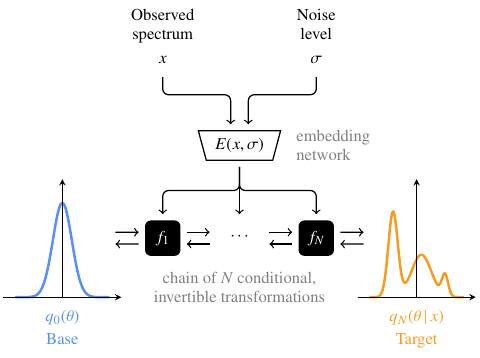}%
            \caption{Discrete Normalizing Flow (DNF)}
        \end{subcaptionblock}%
        \hfill%
        \begin{subcaptionblock}{8.25cm}
            \centering
            \includegraphics{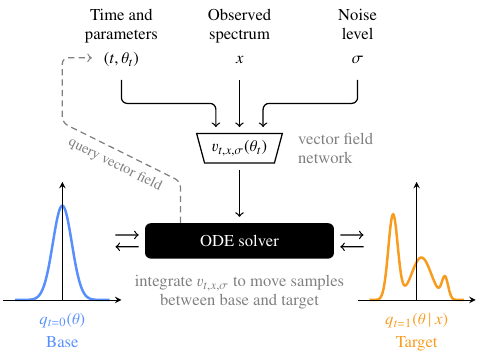}%
            \caption{Continuous Normalizing Flow (CNF)}
        \end{subcaptionblock}%
        \caption{
            Schematic comparison of DNFs and CNFs.
            DNFs (trained with NPE) were first applied to exoplanetary atmospheric retrieval by \citet{Vasist_2023}, while CNFs (trained with FMPE) and the idea of noise level-conditioning are discussed in this work.
        }
        \label{fig:npe-vs-fmpe}
    \end{figure*}

    \begin{figure*}[t]
        \placeonpage{4}
        \centering%
        \includegraphics[]{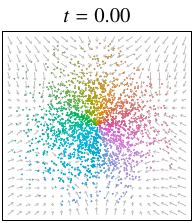}\hfill%
        \includegraphics[]{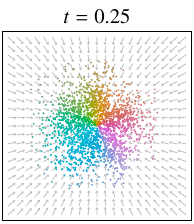}\hfill%
        \includegraphics[]{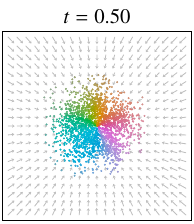}\hfill%
        \includegraphics[]{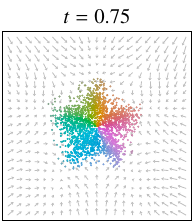}\hfill%
        \includegraphics[]{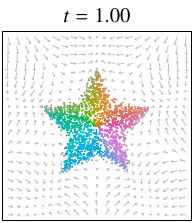}%
        \caption{
            Illustration of a time-dependent vector field (indicated by the gray arrows) continuously transforming samples from a standard 2D Gaussian at $t=0$ into samples from a more complex, star-shaped distribution at $t=1$.
            For simplicity, we show an unconditional example here; for an atmospheric retrieval, the vector field would not only depend on $t$ but also on the observed spectrum~$x$ and the assumed noise level~$\sigma$.
            The size of the arrows has been rescaled for visual purposes.
        }
        \label{fig:vectorfield}
    \end{figure*}

    \begin{figure*}[t]
        \placeonpage{6}
        \centering
        \includegraphics[]{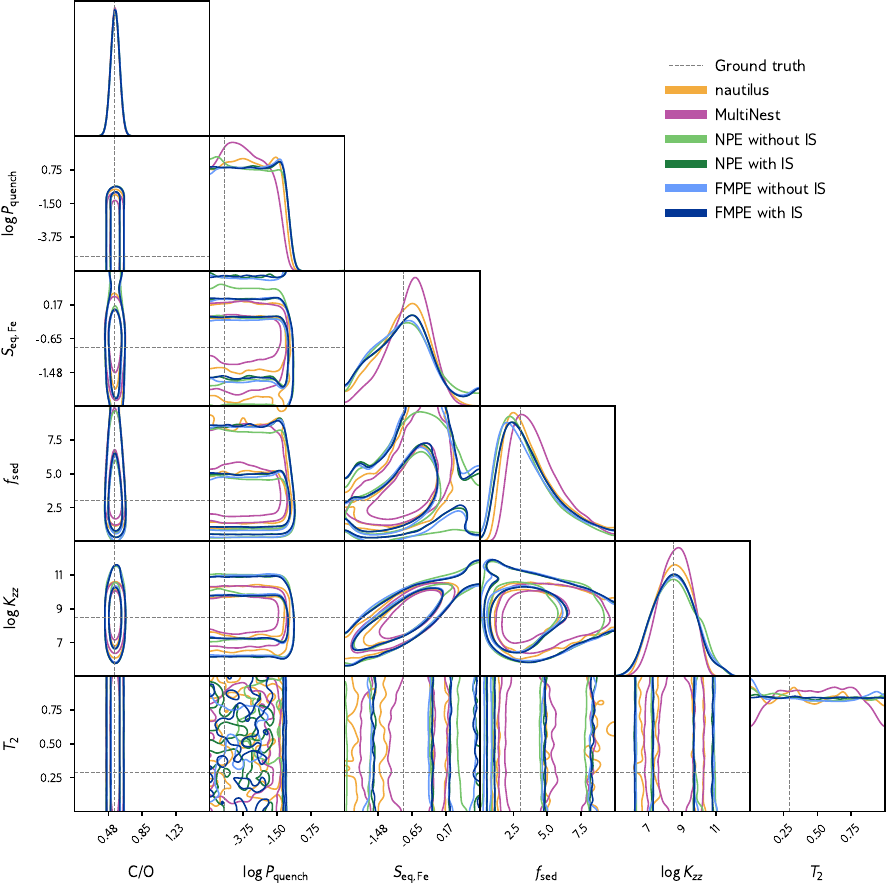}
        \caption{
            Comparison of the 1D and 2D marginal posteriors for the noise-free benchmark spectrum from nested sampling (as implemented by \texttt{nautilus} and \texttt{MultiNest}), FMPE, and NPE.
            For the latter two, we include the results with and without importance sampling.
            For visual purposes, we apply some light Gaussian smoothing to the histograms.
            Furthermore, we only show six selected parameters here; the full version featuring all 16 parameters is found in \cref{fig:cornerplot-full} in the appendix.
        }
        \label{fig:cornerplot-subset}
    \end{figure*}

    \begin{figure*}[t]
        \placeonpage{7}
        \centering
        \begin{subcaptionblock}{4cm}
            \centering
            \includegraphics{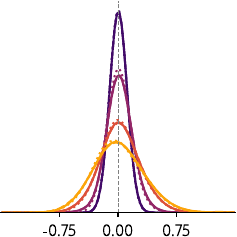}%
            \subcaption{[Fe/H]}
        \end{subcaptionblock}%
        \hfill%
        \begin{subcaptionblock}{4cm}
            \centering
            \includegraphics{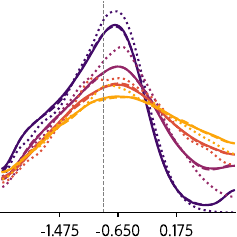}%
            \subcaption{$S_\text{eq,Fe}$}
        \end{subcaptionblock}%
        \hfill%
        \begin{subcaptionblock}{4cm}
            \centering
            \includegraphics{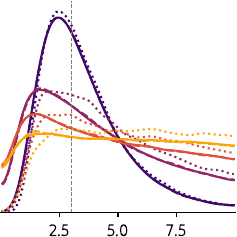}%
            \subcaption{$f_\text{sed}$}
        \end{subcaptionblock}%
        \hfill%
        \begin{subcaptionblock}{4cm}
            \centering
            \includegraphics{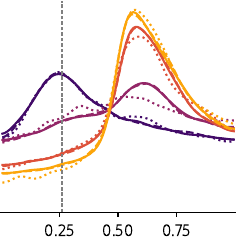}%
            \subcaption{$T_3$}
            \label{subfig:posterior-broadening-d}
        \end{subcaptionblock}%
        {
            \definecolor{C0}{HTML}{3c1364}
            \definecolor{C1}{HTML}{873065}
            \definecolor{C2}{HTML}{cd5a43}
            \definecolor{C3}{HTML}{f0a83d}
            \scriptsize
            \vspace{7pt}
            \hrule\vspace{4pt}
            \textbf{Noise level} (in \qty{1e-16}{\watt \per \meter\squared \per \micro\meter}):\quad
            \mbox{\raisebox{0.4ex}{\protect\tikz{\protect\draw[ultra thick, draw=C0] (0, 0) -- (0.45, 0);}} $\sigma = 0.1$}\quad 
            \mbox{\raisebox{0.4ex}{\protect\tikz{\protect\draw[ultra thick, draw=C1] (0, 0) -- (0.45, 0);}} $\sigma = 0.2$}\quad 
            \mbox{\raisebox{0.4ex}{\protect\tikz{\protect\draw[ultra thick, draw=C2] (0, 0) -- (0.45, 0);}} $\sigma = 0.3$}\quad 
            \mbox{\raisebox{0.4ex}{\protect\tikz{\protect\draw[ultra thick, draw=C3] (0, 0) -- (0.45, 0);}} $\sigma = 0.4$}
            \hfill
            \textbf{Method:}\quad
            \mbox{\raisebox{0.4ex}{\protect\tikz{\protect\draw[ultra thick, solid] (0, 0) -- (0.45, 0);}} FMPE-IS}\quad
            \mbox{\raisebox{0.4ex}{\protect\tikz{\protect\draw[ultra thick, densely dashed] (0, 0) -- (0.45, 0);}} NPE-IS}\quad
            \mbox{\raisebox{0.4ex}{\protect\tikz{\protect\draw[ultra thick, dotted] (0, 0) -- (0.45, 0);}} \texttt{nautilus}}\\
            \vspace{4pt}\hrule
        }
        \caption{
            Marginal posterior distributions for the noise-free benchmark spectrum for four atmospheric parameters at different assumed noise levels~$\sigma$ (encoded by color) and for three different inference methods (encoded by line style).
            In each plot, the $x$-axis spans the prior range, the dashed gray \mbox{line \raisebox{0.4ex}{\protect\tikz{\protect\draw[thick, dash pattern=on 2pt off 1pt, gray] (0, 0) -- (0.42, 0);}}} marks the respective ground truth value, and the $y$-axis shows the density at an arbitrary scale.
        }
        \label{fig:posterior-broadening}
    \end{figure*}

    \begin{figure*}
        \placeonpage{8}
        \centering
        \begin{subcaptionblock}{4.3cm}
            \centering
            \includegraphics{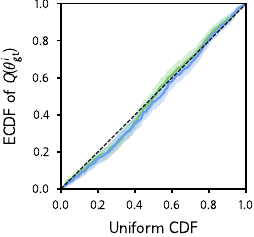}%
            \subcaption{[Fe/H]}
        \end{subcaptionblock}%
        \hfill%
        \begin{subcaptionblock}{4.3cm}
            \centering
            \includegraphics{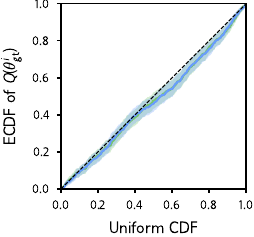}%
            \subcaption{$\log K_{zz}$}
        \end{subcaptionblock}%
        \hfill%
        \begin{subcaptionblock}{4.3cm}
            \centering
            \includegraphics{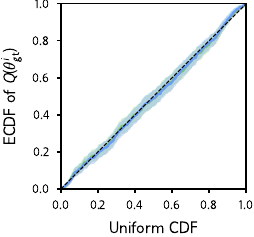}%
            \subcaption{$T_3$}
        \end{subcaptionblock}%
        \hfill%
        \begin{subcaptionblock}{4.3cm}
            \centering
            \includegraphics{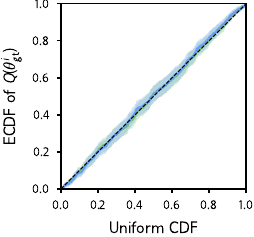}%
            \subcaption{$f_\text{sed}$}
        \end{subcaptionblock}%
        \definecolor{C0}{HTML}{699CFC}
        \definecolor{C1}{HTML}{77C56D}
        \caption{
            P--P plot showing the empirical cumulative distribution functions (ECDFs) of $Q(\theta_\text{gt}^i)$ against a uniform CDF for different atmospheric parameters and for both
            \mbox{\raisebox{0.1ex}{\protect\tikz{\protect\draw[line width=6pt, draw=C0!30] (0, 0) -- (0.45, 0);\protect\draw[line width=1.5pt, draw=C0] (0, 0) -- (0.45, 0);}}}\,\textcolor{C0}{FMPE} and \mbox{\raisebox{0.1ex}{\protect\tikz{\protect\draw[line width=6pt, draw=C1!30] (0, 0) -- (0.45, 0);\protect\draw[line width=1.5pt, , draw=C1] (0, 0) -- (0.45, 0);}}}\,\textcolor{C1}{NPE}.
            The shaded regions indicate the respective \qty{95}{\percent} confidence interval as computed by \texttt{scipy.stats.ecdf()}.
            Panels~(a) and~(b) show the results for the atmospheric parameters with the largest visual deviation from the diagonal (\enquote{worst case}), while panels~(c) and ~(d) show the \enquote{best case.}
            All results are from the default test set.
        }
        \label{fig:pp-plots}
    \end{figure*}

    \begin{figure}[t]
        \placeonpage{8}
        \centering
        \includegraphics{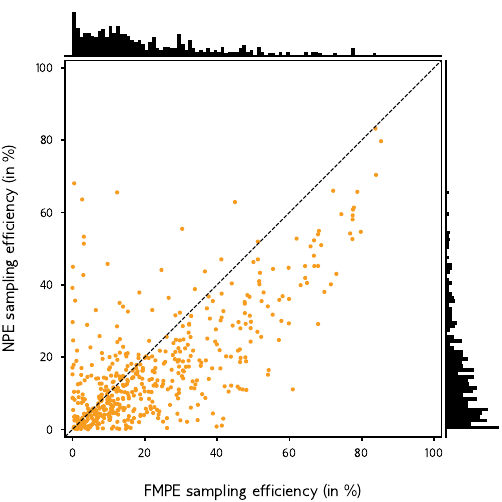}
        \caption{
            Sampling efficiency of FMPE vs. NPE (Gaussian test set).\looseness=-1
        }
        \label{fig:sampling-efficiencies}
    \end{figure}

    \begin{figure*}
        \placeonpage{9}
        \centering
        \includegraphics{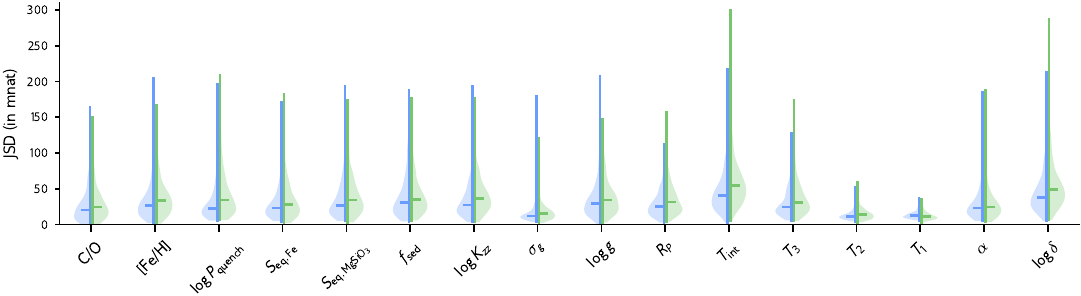}
        \caption{
            \definecolor{C0}{HTML}{699CFC}
            \definecolor{C1}{HTML}{77C56D}
            Distribution of the Jensen-Shannon divergence (JSD) between the posterior marginals with and without IS on the Gaussian test set for both \mbox{\raisebox{0.0ex}{\protect\tikz{\protect\draw[draw=none, fill=C0] (0, 0) rectangle (0.25, 0.25);}} \textcolor{C0}{FMPE}} and \mbox{\raisebox{0.0ex}{\protect\tikz{\protect\draw[draw=none, fill=C1] (0, 0) rectangle (0.25, 0.25);}} \textcolor{C1}{NPE}}.
            Lower is better.
            The vertical line indicates the support of the distribution, while the horizontal bar marks the median.
            We only include retrievals with a sampling efficiency $\varepsilon \geq \qty{1}{\percent}$ to ensure the IS-based posterior is reliable.
        }
        \label{fig:jsd-distribution}
    \end{figure*}

    \begin{table}[t]
        \placeonpage{7}
        \caption{
            Estimate of the Bayesian evidence $\log Z$, the effective sample size ESS, and the sampling efficiency $\varepsilon$ on the benchmark spectrum (with and without noise) for the different methods.
        }
        \label{tab:evidences-benchmark-spectrum}
        \centering
        \begin{subcaptionblock}{\linewidth}
            \caption{Without noise}
            \begin{tabularx}{\linewidth}{
                X %
                c %
                S[table-format = 6.1] %
                S[table-format = 2.1{\,\%}] %
            }
                \toprule
                    Method & 
                    $\log Z$ & 
                    \multicolumn{1}{c}{ESS} &
                    \multicolumn{1}{c}{$\varepsilon$} \\
                \midrule
                    FMPE-IS    & $401.843 \pm 0.002$ & 199761.8 & 19.1\,\% \\
                    NPE-IS     & $401.832 \pm 0.003$ & 121856.5 & 11.6\,\% \\
                    \MultiNest & $407.182 \pm 0.080$ &  30540   & \multicolumn{1}{c}{---} \\
                    \nautilus  & $401.844 \pm 0.010$ &  10004.4 & 3.1\,\% \\
                \bottomrule
            \end{tabularx}
        \end{subcaptionblock}
        \vspace{0pt}
        
        \begin{subcaptionblock}{\linewidth}
            \caption{With noise}
            \begin{tabularx}{\linewidth}{
                X %
                c %
                S[table-format = 6.1] %
                S[table-format = 2.1{\,\%}] %
            }
                \toprule
                    Method & 
                    $\log Z$ & 
                    \multicolumn{1}{c}{ESS} &
                    \multicolumn{1}{c}{$\varepsilon$} \\
                \midrule
                    FMPE-IS    & $198.798 \pm 0.003$ & 111010.8 & 10.6\,\% \\
                    NPE-IS     & $198.792 \pm 0.004$ &  52724.3 & 5.0\,\% \\
                    \MultiNest & $203.235 \pm 0.083$ &  32311   & \multicolumn{1}{c}{---} \\
                    \nautilus  & $198.737 \pm 0.010$ &  10002.9 & 3.2\,\% \\
                \bottomrule
            \end{tabularx}
        \end{subcaptionblock}
        \begin{justify}
            \textbf{Notes:} 
            \MultiNest does not return a value for $\varepsilon$. 
            Therefore, we simply report the default number of equal-weighted posterior samples as its ESS.
            The evidence here is the one without IS; with IS, the $\log Z$ estimates are slightly closer to the other methods.
        \end{justify}
    \end{table}

    \section{Introduction}
    \label{sec:introduction}

    In exoplanet science, the term \enquote{atmospheric retrieval} refers to the inference of atmospheric parameters (e.g., abundances of chemical species, presence of clouds, etc.) from an observed planet spectrum \citep{Madhusudhan_2018}.
    These parameters offer essential insights into the formation and evolution of planets and are central to the study of planetary habitability and the search for life beyond the Solar System.
    Consequently, the atmospheric characterization of exoplanets is also a key driver for the development of future instruments and missions, such as the Habitable Worlds Observatory \citep[HWO;][]{HWO} or the Large Interferometer for Exoplanets \citep[LIFE;][]{Quanz_2022}.
    From a data analysis perspective, the standard approach to atmospheric retrieval is to formulate the task as a Bayesian inference problem.
    In this framework, we start with an observed (emission or transmission) planet spectrum $x \in \mathbb{R}^{b}$ in the form of a vector of flux values in $b$ discrete wavelength bins. 
    The atmospheric parameters of interest are denoted by $\theta \in \Theta$, where for simplicity we can assume $\Theta = \mathbb{R}^d$.
    Any a priori knowledge or belief about the value of $\theta$, including known constraints, is encoded in the prior~$p(\theta)$.
    Moreover, we assume a likelihood function~$p(x \given \theta)$, which describes the probability of observing the spectrum $x$ given some parameter values~$\theta$.
    In practice, the likelihood consists of two parts: 
    (1)~a forward simulator to transform a given $\theta$ into the corresponding spectrum $\hat{x} = \text{simulator}(\theta)$, and (2)~a way to quantify the probability that the true spectrum is $\hat{x}$ when we observe $x$.
    A common approach \citep[see, e.g.,][]{Nasedkin_2024a} is to assume that $x - \hat{x}$ follows a $b$-dimensional Gaussian distribution with mean zero and covariance matrix~$\Sigma \in \mathbb{R}^{b \times b}$.
    This assumption, which yields a simple expression for $p(x \given \theta)$, is equivalent to assuming that the noise in each spectral bin comes from a Gaussian distribution.
    The likelihood and prior together define the Bayesian posterior distribution $p(\theta \given x) = p(x \given \theta)~p(\theta)\,/\,p(x)$, which quantifies the belief about the value of~$\theta$ after observing~$x$.
    The normalization constant $p(x)$ is called the evidence.
    The primary goal of Bayesian inference is now to summarize the posterior in terms of samples $\theta\sim p(\theta \given x)$. 
    
    Traditionally, the posterior $p(\theta \given x)$ is sampled with Markov Chain Monte Carlo (MCMC; see, e.g., \citealt{Hogg_2018} for a practical introduction) or nested sampling \citep{Skilling_2006}.
    These methods essentially run the forward simulator in a loop to explore the parameter space.
    The hardness of this problem scales exponentially with the number of atmospheric parameters, and in practice, millions of likelihood evaluations (i.e., simulator calls) may be required to estimate the posterior of a single exoplanet spectrum. 
    Parallelizing this process is not trivial; see, for example, the discussion in \citet{Ashton_2022}.
    As a result, atmospheric retrievals often imply a considerable computational cost even for the simplest forward models, with overall runtimes on the order of days to weeks. 
    In addition, the simulations from an MCMC or nested sampling run typically cannot be reused to analyze another spectrum---or the same spectrum with different error bars---even when assuming the same prior distribution and simulator settings.
    Finally, the traditional setup is limited to tractable likelihoods where we can evaluate $p(x \given \theta)$ explicitly.  
    This can be restrictive if we want to go beyond the Gaussianity assumption and use realistic instrumental noise for which no analytical expression may be available.

    \paragraph{Related work}
    Due to the computational limitations of traditional methods, alternative approaches have recently received increasing attention---especially those using machine learning~(ML).
    The primary advantage of ML lies in its potential for amortization: while data generation and training may be expensive, the resulting ML model can be used for cheap inference for any number of retrievals without retraining. This amortizes the computational cost of training over inference runs.
    Approaches that fit into this category include deep belief networks \citep{Waldmann_2016}, generative adversarial networks \citep{Zingales_2018}, random forests \citep{MarquezNeila_2018, Fisher_2020, Nixon_2020}, Monte Carlo dropout \citep{Soboczenski_2018}, Bayesian neural networks \citep{Cobb_2019}, and different deep learning architectures \citep{Yip_2021, ArdevolMartinez_2022,Giobergia_2023,Unlu_2023,Sweet_2024}.
    A number of these methods do not produce proper Bayesian posteriors, but rather samples from a \enquote{predictive distribution} \citep{Soboczenski_2018}, whose exact statistical interpretation is not fully understood yet.
    This applies in particular to methods that are trained on noise-free data.
    Approaches that do yield a tractable posterior density commonly rely on simplifying assumptions regarding its form; for example, \citet{Cobb_2019} and \citet{ArdevolMartinez_2022} both assume the posterior to be a multidimensional Gaussian.
    A major advance in this regard is the work of \citet{Vasist_2023}, who bring discrete normalizing flows (DNFs) trained with neural posterior estimation (NPE) to the field of exoplanetary atmospheric retrieval.
    This approach is a simulation-based inference (SBI)%
    \footnote{
        Simulation-based inference (SBI) refers to Bayesian inference based on likelihood simulations $x \sim p(x \given \theta)$. 
        This is in contrast to likelihood-based inference methods (e.g., MCMC) that rely on explicit evaluation of $p(x \given \theta)$.
        For a review of SBI, see, for example, \citet{Cranmer_2020}.
    }
    technique that foregoes explicit assumptions about the functional form of $p(\theta \given x)$ and allows evaluating the posterior density at arbitrary values of $\theta$ instead of only providing sample access.
    DNFs trained with NPE were subsequently used by the winning entry to the 2023 edition of the ARIEL data challenge (\citealt{Aubin_2023}; see also \citealt{Changeat_2023}).
    More recently, the potential of ML has been recognized in non-amortized settings as well:
    \citet{Yip_2024} combine DNFs with variational inference using a differentiable forward simulator, and \citet{ArdevolMartinez_2024} train DNFs using sequential NPE (in particular: automatic posterior transformation as introduced in \citealt{Greenberg_2019}).
    Both approaches do not produce a single model that can be reused for different input spectra, but instead speed up individual retrievals by reducing the number of likelihood evaluations required compared to traditional methods.
    Another class of non-amortized retrieval methods are the approaches of \citet{Himes_2022}, \citet{Hendrix_2023}, \citet{Tahseen_2024}, and \citet{Dahlbuedding_2024}, who do not predict a posterior directly, but instead aim to accelerate retrievals by replacing the computationally expensive simulator with a learned emulator, which can then be used with MCMC or nested sampling.
    In a similar vein, \citet{Gebhard_2023a} propose to learn efficient parameterizations of pressure-temperature profiles to speed up retrievals by reducing the number of required parameters.
    Despite all these advances, significant limitations remain: 
    Although faster than traditional methods, existing ML approaches lack the theoretical guarantees of stochastic samplers, which affects their reliability. 
    In addition, existing ML methods are often trained on noise-free data, or assume a fixed noise level, which limits their practical applicability. 
    These challenges motivate the improvements proposed in this work.

    \paragraph{Contributions}
    We introduce three main ideas to the field of ML-based atmospheric retrieval of exoplanets:%
    \footnote{
        Earlier versions of this work were presented at a AAAI~2024 workshop \citep[see][]{Gebhard_2023b}, AbSciCon~2024, and Exoplanets~5.
    }
    \begin{enumerate*}[label=(\arabic*)]
        \item We adopt flow matching posterior estimation (FMPE; \citealt{Dax_2023b}) with continuous normalizing flows (CNFs) as an alternative to DNFs trained with NPE.
        \item We achieve several advantages compared to existing ML methods for atmospheric retrieval by combining FMPE and NPE with importance sampling (IS). 
        In particular, IS
        \begin{enumerate*}
            \item allows reweighting samples from an ML model towards the true posterior distribution, 
            \item provides the sampling efficiency as a quality metric for the ML-based proposal distribution, and 
            \item offers a way to estimate the Bayesian evidence.
        \end{enumerate*}
        \item We show that we can learn noise level-conditional models that can handle spectra with different assumptions about the size of the error bars and return accurate posteriors without the need to retrain.
    \end{enumerate*}

    \section{Method}
    
    Before presenting our methodical contributions, we briefly summarize NPE using DNFs, which we use as a baseline for CNFs trained with FMPE.
    We note that in \cref{subsec:npe,subsec:fmpe}, we explicitly distinguish between the type of normalizing flow (DNF or CNF) and the training method (NPE or FMPE). 
    In the rest of this paper, we simply write \enquote{NPE} to refer to \enquote{DNFs trained with NPE} and \enquote{FMPE} to refer to \enquote{CNFs trained with FMPE.}
    A schematic comparison of the two methods is found in \cref{fig:npe-vs-fmpe}.

    \subsection{Neural posterior estimation (NPE)}
    \label{subsec:npe}
    
    NPE is an SBI technique introduced by \citet{Papamakarios_2016}.
    The idea is to train a density estimator $q(\theta \given x)$ to approximate the true posterior $p(\theta \given x)$ by minimizing the loss function
    \begin{equation}
        \mathcal{L}_\text{NPE}
        = - \underbrace{\mathbb{E}_{\theta \sim p(\theta)}}_{\substack{\text{sample $\theta$}\\\text{from prior}}}\ \underbrace{\mathbb{E}_{x \sim p(x \given \theta)}}_{\substack{\text{run the\vphantom{p}}\\\text{simulator}}}\ \log q(\theta \given x) \,,
        \label{eq:npe-loss}
    \end{equation}
    where $\mathbb{E}$ denotes the expectation value.
    Unlike in classic supervised learning, this loss function does not require access to the value of the true posterior $p(\theta \given x)$.
    To approximate $p(\theta \given x)$ well, we must choose a sufficiently flexible class of density estimators for $q(\theta \given x)$.
    In practice, $q$ is commonly implemented as a (conditional) discrete normalizing flow (DNF; \citealt{Tabak_2010,Rezende_2015}).
    DNFs construct a bijection between a simple base distribution%
    \footnote{
        It is important to keep in mind that the base distribution $q_0$ of a normalizing flow is different from the prior distribution $p(\theta)$ of a retrieval.
    }
    $q_0$ (e.g., a Gaussian) and a complex target distribution (e.g., a posterior) by applying a chain $\phi_x = f_{N,x}\, \circ\, \cdots\, \circ\, f_{1,x}$ of invertible transformations $f_{i,x}: \mathbb{R}^b \times \Theta \to \Theta$.
    The density of this target distribution is
    \begin{equation}
        q(\theta \given x) 
        = q_0\left( \phi_x^{-1}(\theta) \right) \cdot \left| \det \odv{\phi_x^{-1}(\theta)}{\theta} \right| \,.
        \label{eq:dnf-density}
    \end{equation}
    Each transformation $f_{i,x}$ in this chain is conditional on the observation (indicated by the subscript $x$) and has learnable parameters that are determined by minimizing \cref{eq:npe-loss} during training.
    The Jacobian determinant in \cref{eq:dnf-density} is introduced by the change of variables formula and accounts for the volume changes induced by the transformations $f_{i,x}$.
    It is required to ensure that the learned posterior distribution $q(\theta \given x)$ integrates to~1.
    To keep the evaluation of $q(\theta \given x)$ computationally feasible, we typically have to restrict our choice of $f_{i,x}$ to transformations that yield a triangular Jacobian matrix.
    This results in strong constraints on the architecture of the model, limiting its flexibility and expressiveness.
    For a more detailed introduction to DNFs, see, for example, the reviews by \citet{Kobyzev_2021} or \citet{Papamakarios_2021}.

    \subsection{Flow matching posterior estimation (FMPE)}
    \label{subsec:fmpe}
    
    An alternative approach to density estimation is to parameterize the transformation from the base to the target distribution continuously using a time parameter $t \in [0, 1]$.
    This is the idea of a continuous normalizing flow (CNF; \citealt{Chen_2018a}).
    In the case where the target distribution is a posterior, a CNF consists of a neural network that models a time-dependent vector field $v: \mathbb{R}^{b} \times [0, 1] \times \Theta \to \Theta$.
    This vector field describes the temporal derivative of the trajectory $\phi$ of a sample moving from the base to the target distribution,
    \begin{equation}
        \odv{}{t} \phi_{t, x}(\theta) 
        = v_{t, x}\left( \phi_{t, x}(\theta) \right) \,,
        \label{eq:neural-ode}
    \end{equation}
    where the initial location of the sample under the base distribution is given by $\phi_{t=0, x}(\theta) = \theta_0$.
    Since $x$ is fixed for a given retrieval, we also write $\theta_t \equiv \phi_{t, x}(\theta)$.
    \Cref{eq:neural-ode} is an ordinary differential equation (ODE) where the right-hand side is parameterized by a neural network.
    To obtain samples from the target distribution $q_{t=1, x}(\theta)$, we can draw samples from the base distribution $q_{t=0, x}(\theta) = q_{0}(\theta)$ and transform them by integrating the vector field over time $t$ using a standard ODE solver.
    Intuitively, this process can be interpreted as moving the samples from the base distribution along the time-dependent vector field, which is illustrated in \cref{fig:vectorfield}.
    The main advantage of this continuous approach to density estimation is that we can use an arbitrary neural network to parameterize the vector field $v$: 
    Since we model only the derivative of $v$, there is no need to track a normalization constant.
    This results in greater architectural flexibility, improved scalability to more learnable parameters, and generally faster training \citep{Dax_2023b}.
    On the downside, sampling $q(\theta \given x)$ during inference requires many forward passes through the neural network that parameterizes the vector field $v_{x,t}$ to solve the ODE, instead of just a single forward pass in the case of DNFs.
    
    To obtain an explicit expression for the density $q(\theta \given x)$, we can solve the homogeneous continuity (or transport) equation,
    \begin{equation}
        \pdv{}{t}\, q_{t, x}(\theta) + \operatorname{div}\left( q_{t, x}(\theta) \ v_{t, x}(\theta) \right) = 0 \,.
        \label{eq:continuity-equation}
    \end{equation}
    The continuity equation essentially states that probability is a conserved quantity: if $q_{t=0, x} = q_{0}$ integrates to 1, then so must any $q_{t, x}$, and in particular $q_{t=1, x} = q(\theta \given x)$.
    Solving \cref{eq:continuity-equation} yields
    \begin{equation}
        q(\theta \given x) 
        = q_{1}(\theta_1 \given x)
        = q_{0}(\theta_0) \cdot \exp\left\{ - \int_0^1 \operatorname{div} v_{t, x}(\theta_t) \odif{t} \right\} \,,
        \label{eq:cnf-density}
    \end{equation}
    which describes how the density of the target distribution can be evaluated in terms of the base distribution and the vector field.
    If we want to sample from the target distribution and evaluate the corresponding probability density, we can solve a joint ODE, which is more efficient than sampling and evaluating the density separately; see appendix~C of \citet{Lipman_2022} for details.
    
    So far, we have discussed how to sample and evaluate $q(\theta \given x)$ assuming that we already have a model for the vector field $v$, but not how we can actually learn one.
    In principle, \cref{eq:cnf-density} allows training via likelihood maximization (i.e., NPE), but evaluating the integral over $\operatorname{div} v_{t, x}$ millions or billions of times during training is expensive.
    A much more efficient training scheme for CNFs was proposed by \citet{Lipman_2022}, which was subsequently adapted for Bayesian inference by \citet{Dax_2023b}: flow matching.
    The idea is to train the model for $v$ by regressing it directly onto a target vector field~$u$.
    The trick is that the target vector field $u_t: [0, 1] \times \Theta \to \Theta$ can be chosen on a sample-conditional basis.%
    \footnote{
        We emphasize that \enquote{sample-conditional} (a term introduced in \citealt{Dax_2023a}) is different from the usual conditioning on the observation $x$.
    }
    Sample-conditional means that the target $u_t$ for a given training sample $x$ depends on the corresponding true parameter value~$\theta_1$.
    \mbox{\citet{Lipman_2022}} show that in this case, there exist some simple choices for $u$ that result in a training objective that is equivalent to likelihood maximization.
    For this work, we use
    \begin{equation}
        u_t\left(\theta \mid \theta_1\right)
        = \frac{\theta_1-\left(1-\sigma_\text{min}\right) \theta}{1-\left(1-\sigma_\text{min}\right) t} \,,
        \label{eq:target-vectorfield}
    \end{equation}
    which generates the Gaussian probability paths
    \begin{equation}
        p_t(\theta \given \theta_1) 
        = \mathcal{N}\left( 
            \theta_1\,t,\  
            \left( 1 - (1 - \sigma_\text{min})\,t \right)^2 \cdot I_d
        \right) \,.
    \end{equation}
    This probability path transforms a $d$-dimensional standard normal distribution at $t=0$ into a Gaussian with standard deviation $\sigma_\text{min}$ centered on $\theta_1$ at $t=1$.
    When averaging over many $\theta_1 \sim p(\theta)$, regressing onto these target vector fields $u_t\left(\theta \given \theta_1\right)$ results in a vector field $v_{t,x}(\theta)$ that transforms the base into the correct target distribution.
    Thus, the loss function we need to minimize is
    \begin{equation}
        \mathcal{L}_\text{FMPE}
        = \underbrace{\mathbb{E}_{t \sim p(t)}}_{\substack{\text{sample}\\\text{time $t$}}}
          \underbrace{\mathbb{E}_{\theta_1 \sim p(\theta)}}_{\substack{\text{sample target}\\\text{$\theta_1$ from prior}}}
          \underbrace{\mathbb{E}_{x \sim p(x \given \theta_1)}}_{\substack{\text{run the\vphantom{p}}\\\text{simulator}}}\,
          \underbrace{\mathbb{E}_{\theta_t \sim p_t\left(\theta_t \given \theta_1\right)}}_{\substack{\text{sample $\theta_t$ from}\\\text{probability path}}}
          \left\| r \right\|^2 \,,
        \label{eq:fmpe-loss}
    \end{equation}
    where $r$ is shorthand for the residual vector field
    \begin{equation*}
        r 
        = r(t, \theta_1, x, \theta_t)
        \equiv v_{t, x}( \theta_t ) - u_t( \theta_t \given \theta_1 ) \,.
    \end{equation*}
    \citet{Dax_2023b} provide both theoretical and empirical arguments that this training objective results in posterior estimates that are mass-covering.
    This means that $q(\theta \given x)$ always places some probability mass in regions where the true posterior $p(\theta \given x)$ is non-zero, thus yielding conservative estimates, similar to NPE.

    \subsection{Extension to noise level-conditional models}
    \label{subsec:noise-level-conditional-models}
    
    Previous ML approaches to atmospheric retrieval typically train their models using a fixed noise distribution.
    For example, \citet{Vasist_2023} choose a spectral covariance matrix $\Sigma \in \mathbb{R}^{b \times b}$, with $\Sigma_{ij} = \delta_{ij} \sigma^2$ (i.e., the noise in different bins is uncorrelated), and a fixed value for $\sigma$.
    During training, they sample noise realizations $n \sim \mathcal{N}(0, \Sigma)$ on the fly and add them to the simulated spectra in their training set.
    As a result, at inference time, all retrievals assume a likelihood that is determined by the~$\Sigma$ used during training. 
    To perform an atmospheric retrieval with a different assumption for the noise distribution (e.g., during instrument design), the model must be retrained.
    We overcome this limitation by constructing models that are not only conditioned on the observed spectrum~$x$, but also receive as input a description of the assumed noise distribution that can be varied at inference time.
    One natural choice for such a description is the covariance matrix~$\Sigma$.
    For our proof of concept, we limit ourselves to Gaussian noise with $\Sigma = \sigma^2 \cdot I_d$, with variable~$\sigma$. 
    In this case, the assumed noise distribution is fully described by a single number, the noise level $\sigma$, giving rise to the notion of noise level-conditional models.
    However, extending our approach is straightforward:
    For example, future work may obtain an empirical estimate of the spectral covariance structure using a Gaussian Process \citep[see][]{Greco_2016} and condition on the kernel parameters.
    This would allow letting the assumed noise level vary as a function of wavelength and include correlations between spectral bins.
    
    To learn a noise level-conditional model in practice, we modify the respective neural networks to take not only a spectrum $x$ as input, but also an assumed noise level~$\sigma$ (cf. \cref{fig:npe-vs-fmpe}).
    The updated training procedure for each sample looks as follows:
    \begin{enumerate}
        \item Sample $\theta$ from the prior $p(\theta)$ and simulate the corresponding spectrum $x$, or draw $(\theta, x)$ from a pre-generated training set.
        \item New: Sample $\Sigma$ from a hyper-prior, $\Sigma \sim p(\Sigma)$.
        \item Sample $n \sim \mathcal{N}\left(0, \Sigma\right)$ and add it to the spectrum: $x' = x + n$.
        \item Use $(\theta, x', \Sigma)$ to train the model.
    \end{enumerate}
    This procedure yields a model that can perform multiple retrievals for different assumed error bars on the spectrum without retraining.
    The feasibility of such a conditional approach has previously been demonstrated, for example, by \citet{Dax_2021}, with further refinements by \citet{Wildberger_2023}. 
    In their work, they apply an NPE-based approach to gravitational wave parameter inference and learn a model that can be conditioned on an estimate of the power spectral density (PSD), which describes the detector noise properties at the time of the event being analyzed.

    \subsection{Importance sampling (IS)}
    \label{subsec:importance-sampling}
    
    In practice, both NPE and FMPE can produce results that deviate from the true posterior.
    Using too small a network, too little training data, suboptimal hyperparameter choices, or not training to convergence are just some of the reasons why the predicted posterior $q(\theta \given x)$ may differ from the true posterior $p(\theta \given x)$.
    One way to improve the quality of the predicted posterior post hoc is importance sampling (IS; see, e.g., \citealt{Tokdar_2009} for a review).
    IS is an established statistical procedure that is also used by nested sampling implementations such as \MultiNest \citep{Feroz_2019} or \texttt{nautilus} \citep{Lange_2023}.
    To combine it with ML, the model must allow evaluating the density $q(\theta \given x)$ for arbitrary values of~$\theta$.
    As discussed, this is the case for both FMPE and NPE.
    IS is not limited to amortized approaches, either: for example, \citet{Zhang_2023} have recently described how the idea can be applied to sequential NPE \citep[cf.][]{ArdevolMartinez_2024}.
    
    To improve FMPE or NPE results with IS, we proceed as follows \citep[see][]{Dax_2023a}.
    Assume that we have drawn $N$ samples $\theta_{i} \sim q(\theta \given x)$ from the posterior estimate produced by an ML model.
    We can improve the quality of these samples (in the sense of moving their distribution closer to the true posterior) by reweighting them according to the ratio of the true posterior $p(\theta \given x)$ and the \enquote{proposal} distribution $q(\theta \given x)$.
    To this end, we assign each sample $\theta_i$ an importance weight $w_i$ such that
    \begin{equation*}
        w_i \propto p(\theta_i \given x)\, /\, q(\theta_i \given x) \,.
    \end{equation*}
    Values $\theta_i$ that are more likely under the true posterior than under the proposal thus receive a high weight, while samples that are less likely under $p(\theta \given x)$ than under $q(\theta \given x)$ receive a low weight.
    Moreover, for a perfect proposal distribution, where $p(\theta \given x) = q(\theta \given x)$, all weights are $1$.
    We cannot evaluate $p(\theta \given x)$ directly, but by virtue of Bayes' theorem, we know that $p(\theta \given x) \propto p(x \given \theta) \, p(\theta)$.
    Therefore, we can define the weights as
    \begin{equation}
        w_i 
        = p(x \given \theta_i) \, p(\theta_i)\, /\, q(\theta_i \given x) \,,
        \label{eq:is-weights}
    \end{equation}
    and normalize them such that $\sum_{i=1}^N w_i = N$.%
    \footnote{
        For the practical implementation, it is numerically much more stable to perform these calculations in log-space; see our code for details.
    }
    \Cref{eq:is-weights} also reveals the downside of IS: Computing the weights $w_i$ requires additional likelihood evaluations.
    While these can be fully parallelized, $p(x \given \theta)$ must still be tractable.
    At this point, we can also see why it is crucial that $q(\theta \given x)$ must be mass-covering: 
    Suppose the true posterior has a mode that is not covered by the proposal distribution, that is, the parameter space has a region where $q(\theta \given x)$ is zero but $p(\theta \given x)$ is not.
    In this case, reweighting will not result in samples that are closer to the true posterior distribution because values of $\theta$ that are never sampled cannot be up-weighted.
    
    Using the weights, we define the effective sample size (ESS),
    \begin{equation}
        N_\text{eff} 
        = \left[ \sum_{i=1}^N w_i \right]^2 \Big/ \left[ \sum_{i=1}^N w_i^2 \right] 
        = \frac{N}{\operatorname{var}(w_i) + 1} \,,
    \end{equation}
    and from it the sampling efficiency $\varepsilon$,
    \begin{equation}
        \varepsilon 
        = N_\text{eff} / N \in [0, 1] \,.
    \end{equation}
    The sampling efficiency $\varepsilon$ is a useful diagnostic criterion to assess the quality of the proposal distribution.
    A high efficiency implies that $q(\theta \given x)$ matches $p(\theta \given x)$ well.
    The inverse is not necessarily true, however: 
    $\varepsilon$ is sensitive to even small deviations in just one dimension, meaning that a low efficiency does not automatically imply that $q(\theta \given x)$ is a bad approximation of $p(\theta \given x)$.
    In practice, $\varepsilon \gtrsim \qty{1}{\percent}$ is often already considered a \enquote{good} value.
    
    Finally, importance sampling allows us to estimate the Bayesian evidence $Z$ associated with a posterior. 
    The evidence (also known as the marginal likelihood) plays an important role when comparing models via the Bayes factor, and is defined as
    \begin{equation*}
        Z = \int p(\theta)\ p(x \given \theta) \odif{\theta} \,.
    \end{equation*}
    Using the (unnormalized) weights $w_i$, we can estimate $Z$ as
    \begin{equation}
        Z 
        = \int \frac{q(\theta \given x)\ p(\theta)\ p(x \given \theta)}{q(\theta \given x)} \odif{\theta}
        \approx \frac{1}{N} \sum_{i=1}^N \underbrace{\frac{p(\theta_i)\ p(x \given \theta_i)}{q(\theta_i \given x)}}_{= w_i} \,.
        \label{eq:log-z-estimate}
    \end{equation}
    The standard deviation of $\log Z$ is estimated as:
    \begin{equation}
        \sigma\left( \log Z \right)
        = \sqrt{(N - N_\text{eff})\, /\, (N \times N_\text{eff}) } 
        = \sqrt{(1 - \varepsilon) / (N \varepsilon)} \,.
        \label{eq:sigma-log-z-estimate}
    \end{equation}
    A derivation of \cref{eq:sigma-log-z-estimate} is found, for example, in the supplementary material of \citet{Dax_2023a}, who show the effectiveness of combining NPE with IS for another challenging astronomical parameter inference task, namely gravitational wave analysis.

    \section{Retrieval setup and training dataset}
    \label{sec:simulator-setup}
    
    We test the applicability of our approach using simulated data.
    To maintain comparability with the existing literature, we use a setup very similar to that of \citet{Vasist_2023}, which in turn is based on a study of HR8799~e by \citet{Molliere_2020}.
    We refer to these papers for a more detailed discussion of the atmospheric model.
    For the forward simulator, we use version 2.6.7 of \texttt{petitRADTRANS} \citep{Molliere_2019} with the convenience wrapper from \citet{Vasist_2023}.
    The dimensionality of the atmospheric parameter space, over which we define a uniform prior, is $\dim(\Theta) = 16$; see \cref{app:parameters-priors} for details.
    The simulator maps each $\theta \in \Theta$ to a simulated emission spectrum of a gas giant-type exoplanet. 
    Each spectrum $x$ has a spectral resolution of $R = \lambda / \Delta\lambda = \num{400}$, covering a wavelength range of \qtyrange{0.95}{2.45}{\micro\meter} in $\dim(x) = \num{379}$ bins.
    We also experimented with a resolution of $R=1000$, but found this to be computationally infeasible for the nested sampling baselines.
    A plot with an example spectrum is found in \cref{fig:benchmark-spectrum}.
    
    We generate $2^{25}$ = \num{33554432} spectra as a training dataset for our ML models (plus $2^{20} =$ \num{1048576} spectra for validation) by randomly sampling $\theta$ from the prior and passing it to the simulator (see also \cref{app:effect-of-training-set-size} for an ablation study on the effect of the number of training samples).
    The total upfront cost for this is about \num{12500} CPU hours, which can be fully parallelized.
    The noise is generated on the fly during training using the procedure from \cref{subsec:noise-level-conditional-models}.
    In particular, we sample noise realizations $n$ as
    \begin{equation*}
        n \sim \mathcal{N}\left( 0,\ \sigma^2 \, I_d \right)
        \quad\text{with}\quad
        \sigma \sim \mathcal{U}(0.05, 0.50)\,,
    \end{equation*}
    and pass both $x' = x + n$ (i.e., the noisy spectrum) and $\sigma$ (i.e., the noise level) to the model.
    The unit of $\sigma$ here and in the rest of the paper is \qty{1e-16}{\watt \per \meter\squared \per \micro\meter}.
    The range of the hyper-prior for $\sigma$ is chosen somewhat arbitrarily to cover the (fixed) value of $\sigma = \qty{0.125754e-16}{\watt \per \meter\squared \per \micro\meter}$ used by \citet{Vasist_2023}.%
    \footnote{
        We have confirmed with the authors of \citet{Vasist_2023} that the value of $\sigma = \qty{0.125754e-17}{\watt \per \meter\squared \per \micro\meter}$ in the journal version of their paper was a typographical error.
    }

    \section{Experiments and results}
    \label{sec:experiments}
    
    We train an FMPE and an NPE model with approximately 310 million learnable parameters using stochastic gradient descent.
    Details about the training procedure and the network architectures are found in \cref{app:model-training}.
    Training to convergence on an NVIDIA~H100 GPU takes about 55~hours for the FMPE model, and about 163~hours for the NPE model.
    As a baseline, we also run a number of traditional atmospheric retrievals using two different nested sampling implementations, namely \nautilus and (\texttt{Py})\MultiNest; see \cref{app:nested-sampling} for details.

    \subsection{Experiments on benchmark spectrum}
    \label{subsec:experiments-benchmark-spectrum}
    
    \paragraph{Comparison with nested sampling}
    As a first test case, we use all methods to perform a retrieval of the benchmark spectrum from \citet{Vasist_2023}; see \cref{app:parameters-priors} for details.
    For the noise level, we assume \mbox{$\sigma = 0.125754$}.
    We run the retrieval twice with each method, once without adding any noise to the simulated spectrum, and once with noise randomly drawn from $\mathcal{N}(0, \sigma^2)$.%
    \footnote{
        We remind the reader that there is a difference between the actual noise in the observation $x$ and the noise assumption in the likelihood $p(x \given \theta)$.
    }
    For FMPE and NPE, we use IS with $2^{20} = \num{1048576}$ proposal samples each.
    The main results of these retrievals are shown as a corner plot in \cref{fig:cornerplot-subset} (subset of 6 parameters) and \cref{fig:cornerplot-full} (all 16 parameters).
    We only include the plots for the noise-free retrievals, but we report that the results with noise look very similar and share all the relevant characteristics.
    The corresponding evidences, effective sample sizes, and sampling efficiencies are summarized in \cref{tab:evidences-benchmark-spectrum}.
    
    First, we observe that for well-constrained parameters (such as the C/O ratio or the metallicity), all methods are in excellent agreement.
    Second, without IS, the posteriors from FMPE and NPE show slight deviations for some parameters, while with IS, the results are virtually indistinguishable.
    Given the high sampling efficiencies and large effective sample sizes (see \cref{tab:evidences-benchmark-spectrum}), we are confident that the IS-based results are a very good approximation of the true posterior.
    This is supported by the fact that the posteriors for FMPE and NPE also agree well with the nested sampling results from \nautilus.
    For \MultiNest, we find clear deviations from all other methods.
    These differences are particularly prominent for the quench pressure and the cloud scaling parameters.
    If we visually compare these results with Figure~2 from \citet{Vasist_2023}, we find that the deviations between \MultiNest and the ML-based methods follow the same patterns.
    This is in line with previous findings that \MultiNest seems to have a tendency to produce overconfident posteriors \citep[cf., e.g.,][]{ArdevolMartinez_2022, Dittmann_2024}.
    Fourth, we observe that there are a few instances where \nautilus and \MultiNest place no probability mass on a region of the parameter space where both ML methods agree that the posterior should be non-zero.
    One example of this effect are values of $\log K_{zz} > 11$.
    If the true posterior were indeed zero, as suggested by the nested sampling baselines, we would expect that all ML-based proposal samples in this region would be strongly down-weighted when applying IS.
    The fact that this is not the case leads us to believe that \nautilus and \MultiNest are slightly overconfident.
    This also explains why the peak around the true value is higher for these methods, since the probability mass that should have been placed at $\log K_{zz} > 11$ must be placed elsewhere.
    Finally, if we go beyond the corner plot and compare the log-evidences (see \cref{tab:evidences-benchmark-spectrum}), we find good agreement between FMPE-IS, NPE-IS and \nautilus, while \MultiNest again stands out and deviates more significantly.

    \paragraph{Noise-conditional results}
    To demonstrate the effect of the assumed error bars on the posterior, we repeat the experiment from the last section using four different noise level assumptions: $\sigma = \{0.1, 0.2, 0.3, 0.4\}$.
    For simplicity, we limit ourselves to noise-free retrievals.
    Additionally, we use \nautilus to create a nested sampling baseline for each case.
    We also ran \MultiNest, which confirmed all the trends reported below, but as before deviated clearly from all the other methods.
    Therefore, we do not include it in the selection of the 1D~marginal posteriors shown in \cref{fig:posterior-broadening}:
    First, the results from FMPE and NPE with IS are again virtually indistinguishable, for all parameters and all assumed noise levels.
    Second, the agreement with the results from \nautilus is generally good, with the same caveats as before --- \nautilus seems slightly overconfident in a few cases.
    Further analysis reveals evidence that this might be due to \nautilus missing one or multiple posterior modes.
    Finally, we can see how the assumed noise level $\sigma$ affects the resulting posterior:
    For some parameters, such as the metallicity, the 1D~marginal posterior simply becomes broader as the size of the error bars is increased.
    This makes intuitive sense, as wider error bars mean that we should accept a broader range of potential explanations (i.e., parameter value combinations) for our observation.
    However, some parameters also show non-trivial changes in the posterior:
    For example, in \cref{subfig:posterior-broadening-d}, the posterior does not become broader as we increase $\sigma$, but rather the mode of the distribution shifts away from the ground truth value.
    This effect is observed for all three methods, which suggests that this is the true behavior of the posterior and not a figment of our noise-conditional models.
    Our experiment shows how sensitive the result of an atmospheric retrieval can be to the noise level that one assumes in the likelihood function.
    This illustrates the value of learning an ML model that amortizes over different noise levels, as it allows repeating retrievals quickly and cheaply for different assumptions about the error bars to test the robustness of the values (or bounds) of the inferred atmospheric parameters.

    \paragraph{Sampling efficiency as a diagnostic criterion}
    In \cref{subsec:importance-sampling}, we argued that low values of the sampling efficiency $\varepsilon$ can be used as a diagnostic criterion to identify cases where we might not want to trust the results of our ML model without additional analysis.
    To illustrate this, we run the following simple experiment:
    First, we perform an atmospheric retrieval on a version of the benchmark spectrum to which we have added random Gaussian noise with $\sigma = 0.1$.
    However, to the ML model, we pass a noise level of $\sigma = 0.5$. 
    This corresponds to a conservative retrieval where the error bars are overestimated.
    Then, we repeat the experiment, but swap the values of $\sigma$: We now add noise with $\sigma = 0.5$ to the spectrum but run the retrieval assuming $\sigma = 0.1$.
    This is a misspecification that could yield unreliable results.
    Comparing the results from the two retrievals using our FMPE model with IS (\num{100 000} proposal samples), we find that for the conservative retrieval, the sampling efficiency is $\varepsilon = \qty{22.58}{\percent}$, while for the underestimated noise level, we get $\varepsilon = \num{1.53e-5}$.
    Thus, in this case, the efficiency $\varepsilon$ has indeed flagged the problematic retrieval.

    \subsection{Large-scale analysis}
    \label{subsec:large-scale-analysis}
    
    To confirm that the promising results from \cref{subsec:experiments-benchmark-spectrum} do not only hold for the benchmark spectrum, we perform a large number of retrievals to study the properties of our method for \num{500} different values of $\theta$.
    To this end, we create two test sets: a \enquote{default} one, where we sample $\theta$ directly from the Bayesian prior, and a \enquote{Gaussian} one, consisting of spectra from a neighborhood around the parameters $\theta_0$ of the benchmark spectrum.
    The exact details and motivation for using two test sets are described in \cref{app:subsec:test-set-generation}.

    \paragraph{Faithfulness of posteriors}
    It is important for the interpretability of an ML-based posterior that the estimated distribution is well-calibrated, that is, neither broader (i.e., underconfident) nor tighter (i.e., overconfident) than the true posterior.
    To probe this for our models, we perform a retrieval for each spectrum in the default test set, drawing \num{100 000} posterior samples each.
    In the following, we consider every atmospheric parameter~$\theta_i$ separately (where $\theta_i$ is the $i$-th entry of the parameter vector $\theta$).
    For each retrieval, we determine the quantile~$Q(\theta_{i}^\text{gt})$ of the ground-truth value~$\theta^{\mathrm{gt}}_i$.
    In other words, for a given retrieval, $Q(\theta_i^\text{gt})$ is the fraction of posterior samples where the value of $\theta_i$ is less than the ground truth.
    If the posterior estimates are well-calibrated, we expect that these quantiles~$Q(\theta_i^\text{gt})$ should follow a uniform distribution:
    For example, the ground-truth value should fall into the lower half of the posterior exactly \qty{50}{\percent} of the time.
    To test this, we create a P--P~plot, where we plot the empirical cumulative density function (ECDF) of the $Q(\theta_i^\text{gt})$ values against the CDF of a uniform distribution.
    This should yield a diagonal line (up to statistical deviations).
    If the posteriors are underconfident, values of $Q(\theta_i^\text{gt})$ close to 0.5 are overrepresented compared to values near 0 or 1, and the resulting shape of the P--P~plot looks like a sigmoid function:~\raisebox{-2pt}{\protect\tikz{\protect\draw[very thin, gray] (-0.15, -0.15) -- (0.15, 0.15); \draw[semithick] (-0.15, -0.15) edge[in=180, out=0] (0.15, 0.15);}}.
    Conversely, for overconfident posteriors, we expect a shape that resembles a $\sinh$-function:~\raisebox{-2pt}{\protect\tikz{\protect\draw[very thin, gray] (-0.15, -0.15) -- (0.15, 0.15); \draw[semithick] (-0.15, -0.15) edge[in=270, out=90] (0.15, 0.15);}}.
    A representative selection of our results is shown in \cref{fig:pp-plots}.
    The P--P~plots look well-calibrated for all atmospheric parameters, with no clear signs of under- or overconfidence.
    The differences between FMPE and NPE are negligible.
    We note that a similar experiment with nested sampling is computationally infeasible: 
    From the runtimes listed in \cref{tab:nested-sampling-stats}, we can estimate that running 500 traditional retrievals would easily require about over a million CPU hours.

    \paragraph{Sampling efficiency and proposal quality}
    We repeat the experiment from the last section using the Gaussian test set, that is, we perform 500 retrievals for both the FMPE and the NPE model drawing \num{100 000} proposal samples each and applying IS.%
    \footnote{
        All results in this section hold qualitatively also for the default test set.
    }
    In \cref{fig:sampling-efficiencies}, we compare the corresponding sampling efficiencies for FMPE and NPE.
    First, there is a clear correlation ($\rho = 0.696$) between the two methods.
    This could be explained by the fact that a key challenge for both FMPE and NPE is the extraction of relevant information from the given spectra, which is independent of the actual density estimation task and equally difficult for both methods.
    Second, FMPE-IS yields on average a higher sampling efficiency than NPE-IS.
    This indicates that the posterior estimates from FMPE are closer to the true posterior than those from NPE.
    To confirm this, we look at the Jensen-Shannon divergence (JSD) between the marginal posteriors for each atmospheric parameter with and without IS.
    The JSD is a measure of the similarity of distributions: the lower the JSD, the more similar the distributions are. 
    Looking at the distribution of the JSD across the test set (\cref{fig:jsd-distribution}), the average difference between the FMPE and FMPE-IS posteriors is smaller than the difference between the NPE and NPE-IS posteriors for all but one parameter ($T_1$), which supports the above hypothesis.
    The fact that $T_1$ yields the lowest JSD value is likely not a coincidence: 
    $T_1$ describes the temperature at the very upper end of the atmosphere.
    This property is notoriously difficult to constrain from emission spectra (cf., e.g., Sect.~4.4 of \citealt{Gebhard_2023a}), and consequently the resulting posterior for $T_1$ is, in most cases, equal to its prior.

    \paragraph{Bayesian evidence}
    Finally, we look at the estimates for the Bayesian (log)-evidence $\log Z$ produced by FMPE-IS and NPE-IS.
    Due to the prohibitive computational cost of running 500 retrievals with nested sampling, we cannot compare these values to those from traditional methods.
    However, we can report that in more than half of the cases, the $\log Z$ estimates from FMPE-IS and NPE-IS agree within one standard deviation, and in \qty{98}{\percent} of the cases, the agreement is within three standard deviations.

    \subsection{Runtime comparison at inference time}
    
    Finally, we compare the runtimes of FMPE and NPE at inference time.
    For NPE, drawing $2^{16} = \num{65536}$ posterior samples takes \qty{1.161 +- 0.004}{\second}, while sampling and evaluating log-probability densities takes \qty{1.163 +- 0.002}{\second}.
    In contrast, the FMPE model requires \qty{16.262 +- 1.433}{\second} for sampling and \qty{258.08 +- 4.93}{\second} for evaluating log-probabilities when using the settings from \cref{app:subsec:inference-procedure}. 
    This significant difference is because NPE only needs one forward pass through the neural network, whereas FMPE requires multiple model evaluations by the ODE solver.
    All runtimes were measured on an NVIDIA~H100 GPU and averaged over 10 runs.

    \section{Discussion}
    
    \paragraph{Amortization}
    Previous work has mainly considered the potential of ML-based methods for amortization in terms of the ability to efficiently process large numbers of different spectra that share similar instrument characteristics, as we can expect from future missions such as ARIEL \citep{Tinetti_2021}.
    In this work, we have opened another perspective: the potential of using amortized models to perform atmospheric retrievals under different assumptions about the noise in the observed spectrum.
    This is useful in several ways: First, even for the same instrument, the expected noise level in a spectrum is not always the same, but may depend, for example, on the host star of the target planet.
    Being able to handle these different scenarios without retraining makes a model significantly more useful in practice and increases the number of cases over which the training costs can be amortized.
    Second, being able to efficiently run retrievals for different noise levels can help to understand how sensitive the inferred parameters and constraints are to these assumptions, and thus make atmospheric retrievals more robust. 
    Finally, amortization across different noise assumptions may become particularly valuable during the design phase of new instruments and missions for exoplanet science, such as HWO or LIFE:
    To understand the effect of specific design choices on the expected yield or the ability to constrain certain exoplanet properties, it is necessary to simulate surveys with different instrument configurations, which is expensive with traditional methods.
    For example, \citet{Konrad_2022} explore the minimum requirements for LIFE and quantify how well various atmospheric parameters can be constrained as a function of the signal-to-noise ratio (S/N).
    Our method would allow performing this type of analysis at a much higher resolution (e.g., 100 S/N values instead of 4) with minimal additional computational cost.

    \paragraph{Importance sampling}
    For a mass-covering posterior estimate $q(\theta \given x)$, which can generally be assumed for FMPE and NPE \citep[see][]{Dax_2023a,Dax_2023b}, IS reweights samples from $q(\theta \given x)$ to follow the true posterior $p(\theta \given x)$ exactly in the asymptotic case.
    Of course, IS is not a magic bullet: If $q$ is a poor approximation of~$p$, the variance of the importance weights is high and the resulting ESS small. 
    In this case, the IS posterior is not reliable.
    Conversely, a high importance sampling efficiency~$\varepsilon$ is a strong indicator that $q(\theta \given x)$ is close to the true posterior.
    Using~$\varepsilon$ as a quality criterion thus allows us to go beyond statistical arguments about the well-calibratedness of our posteriors in the average case and reason quantitatively about the reliability of our model outputs for individual retrievals.
    Furthermore, IS allows us to estimate the Bayesian evidence associated with a retrieval result, something that previous amortized ML approaches have not addressed.
    The evidence plays an important role in model comparison via the Bayes factor, which is given by the evidence ratio.%
    \footnote{
        For details and a note of caution, see, e.g., \citet{Jenkins_2011}. 
    }
    Access to the evidence therefore facilitates further studies, for example, to test if a planet spectrum is better explained by a model with or without clouds, or with or without a particular chemical species.

    \paragraph{Runtime considerations}
    A key difference between FMPE and NPE is the tradeoff between the time required to train a model and the time to do inference:
    Our FMPE model is substantially faster to train than the NPE model (factor of~$\sim\!3$), but is also considerably slower at inference time, namely by a factor of~$\sim\!10$ for sampling, and a factor of~$\sim\!200$ when evaluating log-probability densities.
    Of course, we must not forget that the training time is usually on the order of days, while the inference time is measured in seconds.
    Furthermore, evaluating log-probabilities is mainly needed for IS, in which case we have to consider two more things:
    First, FMPE yields on average higher sampling efficiencies than NPE, meaning that fewer proposal samples are required to achieve the same ESS.
    Second, when running IS, the computational cost of the inference procedure is dominated by the required additional likelihood evaluations, which usually significantly outweigh the cost of drawing proposal samples.
    It is also important to note that both sampling from the proposal and IS can be arbitrarily parallelized, which is a key advantage over nested sampling or MCMC:
    While the true strength of amortized approaches is, of course, in repeated inference, the ability to easily run computations in parallel on hundreds or thousands of CPUs can lead to scenarios where generating training data, training a model, and running inference for only a single retrieval can still be preferable (in terms of wall time) to traditional methods.

    \paragraph{Future directions}
    Going forward, we see multiple directions to build on and extend the ideas we have presented in this work.
    First, while we have shown that models can be trained to amortize over different noise levels, we have not yet considered noise levels that vary as a function of wavelength or, more importantly, correlations of noise across different bins.
    The dangers of ignoring spectral covariance in atmospheric retrievals have been discussed, for example, in \citet{Greco_2016} and \citet{Nasedkin_2023}.
    Extending our approach in this direction is straightforward, but will require close collaboration with the instrumentation community to ensure that the resulting noise models are rooted in reality. 
    Second, our ideas can be transferred to similar problems in adjacent fields, such as inferring parameters of exoplanet interiors, or analyzing metal abundances in polluted white dwarfs, where ML has recently been applied \citep[see, e.g.,][]{Haldemann_2023,Baumeister_2023,BadenasAgusti_2024}.
    Third, our framework could be extended to adaptive priors: 
    Some atmospheric parameters, such as the planet radius and surface gravity, can be constrained before a retrieval using coarse methods, or may be known from other sources (e.g., radial velocity measurements).
    In this case, we might want to use an \enquote{informed prior} \citep[cf.][]{Hayes_2020}.
    Prior conditioning~\citep{Dax_2024} can integrate such information to simplify the inference task, while retaining the amortized properties of our framework.  
    Finally, in this work, we have only considered setups where the target spectrum of a retrieval was generated using the same atmospheric model that produced the training data.
    Future work should investigate in more detail how our approach performs on out-of-distribution data, or in the case of model mismatch, that is, when the assumed forward model is not a good fit for the actual atmosphere (e.g., the simulator assumes no clouds when the planet is cloudy).

    \section{Summary and conclusions}
    
    In this work, we have established FMPE \citep{Dax_2023b} as a new amortized ML approach to atmospheric retrieval. 
    FMPE retains the desirable properties of NPE, such as distributional expressiveness (i.e., no simplifying assumptions about the form of the posterior), tractable posterior density, and simulation-based training, while offering a different trade-off between training and inference time. 
    Due to the lack of architectural constraints on the neural networks, FMPE is also more scalable than NPE. 
    In the amortized scenario (i.e., ignoring training costs), both FMPE and NPE are orders of magnitude faster than traditional samplers; however, their high degree of parallelizability can make them a powerful alternative even for a single retrieval.
    In terms of posterior accuracy, FMPE and NPE are both on par with traditional nested sampling, especially when combining them with importance sampling.
    We demonstrated the value of IS for atmospheric retrieval in several ways: 
    IS not only reweights the proposal samples from an ML model to more closely match the true posterior, it also provides a metric (the sampling efficiency) for when the model outputs can be trusted. 
    Finally, IS produces an accurate estimate of the Bayesian evidence, which is important for model comparison, and which previous amortized ML approaches did not provide.
    Although this comes at an increased computational cost (for additional likelihood evaluations), we believe that these clear benefits justify that IS should become a standard tool for ML-based atmospheric retrieval.
    Finally, we have shown how to learn ML models that are noise level-conditional and enable amortized inference without retraining not only for different spectra, but also for the same spectrum with different error bars. 
    This allows testing the robustness of inferred parameter values with respect to the assumed noise level, and is also highly relevant for design studies of future instruments (e.g., HWO or LIFE).
    Such studies need to analyze the impact of different configurations on the ability to constrain atmospheric parameters, and will clearly benefit from being able to explore, for example, a wide range of signal-to-noise ratios at almost no additional computational cost.

    \vspace{6pt}
    \noindent
    \textbf{Code and data availability:}
    Our entire code is available online at \href{https://github.com/timothygebhard/fm4ar}{github.com/timothygebhard/fm4ar}, and our data and results can be downloaded from \href{https://doi.org/10.17617/3.LYSSVN}{doi.org/10.17617/3.LYSSVN}.

    \begin{acknowledgements}
        The authors thank Kai Hou Yip for the fast and constructive review.
        We also thank Malavika Vasist and Evert Nasedkin for their help with the data generation code, and Jean Hayoz and Emily Garvin for valuable comments on the manuscript.
        Parts of this work have been carried out within the framework of the National Centre of Competence in Research PlanetS supported by the Swiss National Science Foundation under grant \mbox{51NF40\_20560}.
        The computational work for this manuscript was performed on the Atlas cluster at the Max Planck Institute for Intelligent Systems.
        TDG acknowledges funding through the Max Planck ETH Center for Learning Systems.
        MD thanks the Hector Fellow Academy for support.
        The authors thank the International Max Planck Research School for Intelligent Systems (IMPRS-IS) for supporting AK.
        SPQ acknowledges the financial support of the SNSF.
        This work has made use of many open-source Python packages, including
        \texttt{corner} \citep{ForemanMackey_2016},
        \texttt{dynesty} \citep{Speagle_2020},
        \texttt{glasflow} \citep{Williams_2024},
        \texttt{matplotlib} \citep{Hunter_2007},
        \texttt{multiprocess} \citep{McKerns_2012},
        \texttt{nautilus-sampler} \citep{Lange_2023},
        \texttt{normflows} \citep{Stimper_2023},
        \texttt{numpy} \citep{Harris_2020},
        \texttt{scipy} \citep{Virtanen_2020},
        \texttt{scikit-learn} \citep{Pedregosa_2011},
        \texttt{pydantic} \citep{Colvin_2024},
        \texttt{PyMultiNest} \citep{Buchner_2014},
        \texttt{pandas} \citep{McKinney_2010},
        \texttt{petitRADTRANS} \citep{Molliere_2019},
        \texttt{torch} \citep{Paszke_2019}, and
        \texttt{ultranest} \citep{Buchner_2021}.
        The colorblind-friendly color schemes in our figures are based on \citet{Petroff_2021}.
    \end{acknowledgements}

    \bibliographystyle{aa}
    \bibliography{main.bib}

    \appendix

    \section{Atmospheric parameters and priors}
    \label{app:parameters-priors}
    
\begin{figure*}[t]
        \centering
        \includegraphics[]{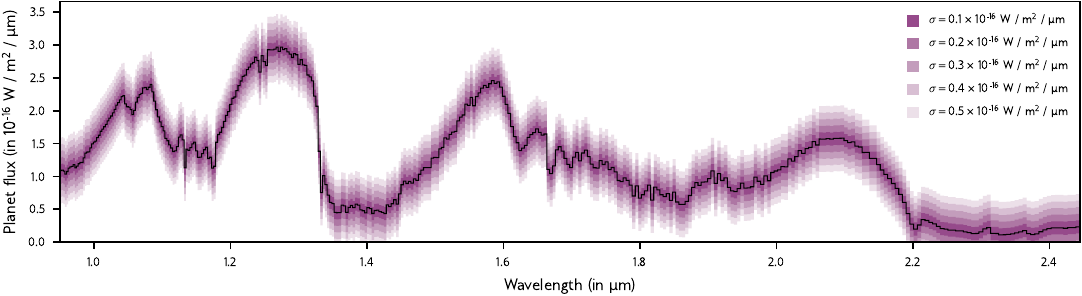}\\
        \caption{
            Noise-free benchmark spectrum (black line) obtained by running the simulator for $\theta_0$ from \cref{tab:atmospheric-parameters}.
            The colored bands illustrate different noise levels up to the largest one used for training our noise level-conditional models (see \cref{sec:simulator-setup}).
        }
        \label{fig:benchmark-spectrum}
    \end{figure*}

    In \cref{tab:atmospheric-parameters}, we provide an overview of the 16 atmospheric parameters we consider (see \cref{sec:simulator-setup}), as well as their priors.
    The table also shows $\theta_0$, that is, the parameter values for the benchmark spectrum that we use in \cref{subsec:experiments-benchmark-spectrum} and that is shown in \cref{fig:benchmark-spectrum}.
    We adopted our choice of parameters and priors from \citet{Vasist_2023} (cf. their Table~1 and Table~2), who in turn based it on \citet{Molliere_2020}.
    
    We note two more things:
    The first comment concerns the name and interpretation of the parameters for the optical depth and the pressure-temperature (PT) profile.
    Tables~1 and~2 of \citet{Vasist_2023} (and Tables~1 and~3 of \citealt{Molliere_2020}) use a different convention than the actual Python implementation, which is the one that we adopt in \cref{tab:atmospheric-parameters}.
    To clarify the differences and prevent any confusion, we explain the conventions and their connection in greater detail.
    The atmospheric model we use describes the PT~profile using four temperature parameters in Kelvin, called $T_0$ through $T_3$ by \citet{Molliere_2020} and \citet{Vasist_2023}, as well as the parameters $\alpha$ and $\delta$, which relate the optical depth~$\tau$ to the atmospheric pressure $P$ via a power law,
    \begin{equation}
        \tau = \delta \cdot P^\alpha \,.
        \label{eq:optical-depth}
    \end{equation}
    The temperature $T_0$, which is called \texttt{T\_int} in the \petitRADTRANS code, describes the interior temperature of the planet and appears as a free parameter in the Eddington approximation used for the photosphere (see eq.~2 in \citealt{Molliere_2020}),
    \begin{equation}
        T(\tau)^4 = \nicefrac{3}{4} \cdot T_0^4 \cdot (\nicefrac{2}{3} + \tau) \,.
        \label{eq:photosphere-temperature}
    \end{equation}
    For $T_1$ to $T_3$, which are the support points of a cubic spline description of the PT profile at high altitudes, the priors defined by \citet{Molliere_2020} and \citet{Vasist_2023} are not independent; for example, the prior assumed for $T_1$ is $\mathcal{U}(0, T_2)$.
    From a physical perspective, this makes sense, as we want to enforce $T_\text{connect} > T_3 > T_2 > T_1$, where $T_\text{connect}$ is the temperature of the uppermost photosphere layer at $\tau = 0.1$; see \cref{eq:photosphere-temperature}.
    However, for sampling, it is more efficient to use a parameterization where we can draw all atmospheric parameters independently.
    Presumably for this reason, the \petitRADTRANS method handling the simulation of the spectra (namely \href{https://gitlab.com/mauricemolli/petitRADTRANS/-/blob/09dbc6d3492c0608bb5adc711d6fc3f178d7d4dd/petitRADTRANS/retrieval/models.py}{\texttt{emission\_model\_diseq()}\extlink}) uses such an independent re-parameterization:
    The function expects four temperature arguments: \texttt{T\_int} in Kelvin, and \texttt{T1}, \texttt{T2}, \texttt{T3} all dimensionless in $[0, 1]$.
    These inputs are then overwritten as
    \begin{align*}
        \texttt{T3} &\leftarrow \sqrt[4]{ \nicefrac{3}{4} \cdot \texttt{T\_int}^4 \cdot (0.1 + \nicefrac{2}{3}) } \cdot (1 - \texttt{T3}) \,, \\
        \texttt{T2} &\leftarrow \texttt{T3} \cdot (1 - \texttt{T2}) \,, \\
        \texttt{T1} &\leftarrow \texttt{T2} \cdot (1 - \texttt{T1}) \,.
    \end{align*}
    The temperature values from Table~2 in \citet{Vasist_2023} can be obtained by inserting the respective $\theta_0$ values from \cref{tab:atmospheric-parameters} into the above conversion formulas.
    Finally, \texttt{emission\_model\_diseq()} expects an input argument \texttt{log\_delta} in $[0, 1]$, which is, however, not simply the logarithm of the $\delta$ parameter in \cref{eq:optical-depth}.
    Instead, \texttt{log\_delta} is used internally to compute $\delta$ via the transformation
    \begin{equation*}
        \delta \leftarrow \left( 10^6 \cdot 10^{-3 + 5 \cdot \texttt{log\_delta}} \right)^{-\alpha} \,.
    \end{equation*}
    As before, plugging the $\theta_0$ values from \cref{tab:atmospheric-parameters} into this transformation and taking the logarithm yields the $\log \delta$ value from Table~2 in \citet{Vasist_2023}.
    We have chosen to report the prior ranges and $\theta_0$ values of the independent re-parameterization in \cref{tab:atmospheric-parameters} (i.e., the inputs to the \texttt{emission\_model\_diseq()} function before internal conversions) to maintain consistency with the code and to use the same prior bounds for all retrievals.
    
    The second comment concerns the parameters that \citet{Vasist_2023} use to describe the cloud mass fraction abundance, namely $\log \tilde{X}_{\mathrm{Fe}}$ and $\log \tilde{X}_{\mathrm{MgSiO}_3}$.
    The meaning of the corresponding arguments to the \texttt{emission\_model\_diseq()} function, \texttt{log\_X\_cb\_Fe(c)} and \texttt{X\_cb\_MgSiO3(c)}, changed in version 2.4.8 of \petitRADTRANS.
    To get the same behavior as before, we use the arguments \texttt{eq\_scaling\_Fe(c)} and \texttt{eq\_scaling\_MgSiO3(c)} instead and label the parameters as $S_\text{eq,Fe}$ and $S_\text{eq,\ch{MgSiO3}}$, respectively.
    The prior ranges and $\theta_0$ values remain unchanged.

    \begin{table*}[t]
        \centering
        \caption{
            Overview of the 16 atmospheric parameters considered in this work, with the respective priors used for all retrievals, as well the parameter values $\theta_0$ of the \enquote{benchmark spectrum} (cf. \cref{subsec:experiments-benchmark-spectrum}).
        }
        \label{tab:atmospheric-parameters}
        \begin{tabularx}{\linewidth}{llS[table-format=4.2]X}
        \toprule
        \multicolumn{1}{l}{\textbf{Parameter}} & \textbf{Prior} & \textbf{$\theta_0$ value} & \textbf{Meaning} \\                                                
        \midrule
        $\ch{C}/\ch{O}$            &  $\mathcal{U}(0.1, 1.6)$   &     0.55  &  Carbon-to-oxygen ratio \\
        $[\ch{Fe}/\ch{H}]$         &  $\mathcal{U}(-1.5, 1.5)$  &     0.00  &  Metallicity \\
        $\log P_\text{quench}$     &  $\mathcal{U}(-6.0, 3.0)$  &    -5.00  &  Pressure below which the \ch{CO}, \ch{CH_4} and \ch{H_2O} abundances are taken as vertically constant \\
        $S_\text{eq,Fe}$           &  $\mathcal{U}(-2.3, 1.0)$  &    -0.86  &  Scaling factor for equilibrium cloud abundances (\ch{Fe}) \\
        $S_\text{eq,\ch{MgSiO3}}$  &  $\mathcal{U}(-2.3, 1.0)$  &    -0.65  &  Scaling factor for equilibrium cloud abundances (\ch{MgSiO_3}) \\
        $f_\text{sed}$             &  $\mathcal{U}(0.0, 10.0)$  &     3.00  &  Sedimentation parameter for both cloud types \\
        $\log K_{zz}$              &  $\mathcal{U}(5.0, 13.0)$  &     8.50  &  Vertical eddy diffusion coefficient (in \unit{\centi\meter\squared\per\second}) \\
        $\sigma_g$                 &  $\mathcal{U}(1.05, 3.0)$  &     2.00  &  Width of the (log-normal) cloud particle size distribution \\
        $\log g$                   &  $\mathcal{U}(2.0, 5.5)$   &     3.75  &  Surface gravity \\
        $R_P$                      &  $\mathcal{U}(0.9, 2.0)$   &     1.00  &  Planet radius (in Jupiter radii) \\
        $T_\text{int}$             &  $\mathcal{U}(300, 2300)$  &  1063.60  &  Interior temperature of the planet (in Kelvin) \\
        $T_3$                      &  $\mathcal{U}(0.0, 1.0)$   &     0.26  & Inner support point for pressure-temperature spline at high altitudes \\
        $T_2$                      &  $\mathcal{U}(0.0, 1.0)$   &     0.29  & Middle support point for pressure-temperature spline at high altitudes \\
        $T_1$                      &  $\mathcal{U}(0.0, 1.0)$   &     0.32  &  Outer support point for pressure-temperature spline at high altitudes \\
        $\alpha$                   &  $\mathcal{U}(1.0, 2.0)$   &     1.39  &  Exponent in $\tau = \delta P^\alpha$ (i.e., the assumed power law for the optical depth $\tau$) \\
        $\log \delta$              &  $\mathcal{U}(0.0, 1.0)$   &     0.48  &  Proportionality factor in $\tau = \delta P^\alpha$ (see above) \\
        \bottomrule
        \end{tabularx}
    \end{table*}
    
    \section{Model and training details}
    \label{app:model-training}
    
    In this appendix, we describe the model configurations and details of the training procedure we used for the experiments in \cref{sec:experiments}.
    For the full technical details, we recommend a look at our code.

    \subsection{Flux and parameter normalization}
    
    To rescale the flux values that are passed to the neural networks, we follow \citet{Vasist_2023} and use a softclip function, where each value $x_i$ is mapped to $x_i \mapsto x_i \, / \, (1 + |x_i/\beta|) \in [-\beta, \beta]$ with bound $\beta = 100$.
    The atmospheric parameters are standardized by subtracting the mean~$\mu$ and dividing by the standard deviation~$\sigma$.
    Both $\mu$ and $\sigma$ are computed from the corresponding prior.

    \subsection{Model configurations}
    
    \paragraph{FMPE model}
    There are three parts to the FMPE model: a context encoder, a $(t, \theta_t)$-encoder, and a network modeling the vector field. 
    All three parts are implemented as deep residual neural networks (\enquote{resnet}) that use layer normalization \citep{Ba_2016}, GELU activation functions \citep{Hendrycks_2016}, and dropout regularization \citep{Hinton_2012} with $p=0.1$.
    The context encoder concatenates the normalized flux values and assumed error bars for each spectral bin and passes the result through a 3-layer resnet with \num{1024} units each to compute the context embedding.
    The $(t, \theta_t)$-encoder first applies a sine\,/\,cosine positional encoding to $t$ before stacking it with the original $(t, \theta_t)$-tuple and passing it through another 3-layer resnet.
    Then, the embedded context and the embedded $(t, \theta_t)$-values are combined using a gated linear unit (GLU; \citealt{Dauphin_2016}) on the latter.
    The result is finally passed to the vector field network, which consists of 18 residual layers, with layer sizes decreasing from \num{8192} to \num{16} (i.e., the number of atmospheric parameters).
    The total number of learnable parameters for this configuration is \num{313 443 376}.
    Our architecture and hyperparameters are based on a number of preliminary experiments to identify a combination that works well; however, a more systematic optimization is left for future work.

    \paragraph{NPE model}
    The model consists of two parts: an encoder to compress the context to a lower-dimensional representation, and the actual DNF.
    For the context encoder, we normalize the flux values (see above) and concatenate them with the error bars before passing them to a 15-layer resnet with layer sizes decreasing from \num{4096} to \num{256}.
    We use the same activation and regularization layers as for the FMPE model.
    The output of the context encoder is passed to each step of the DNF.
    We use a neural spline flow (NSF; \citealt{Durkan_2019}) as implemented by the \texttt{glasflow} library \citep{Williams_2024}, which is itself based on \texttt{nflows} \mbox{\citep{Durkan_2020}}.
    The NSF consists of 16 steps, each consisting of 4 blocks with \num{1024} units.
    The number of bins for the splines is 16.
    This configuration results in a total of \num{310 120 848} learnable parameters.
    As for FMPE, the NPE model may benefit from more hyperparameter optimization, which we leave for future work.\looseness=-1

    \subsection{Training procedure}
    
    We train both our FMPE and our NPE model for 600 epochs using stochastic gradient descent in the form of \texttt{AdamW} \citep{Loshchilov_2017}, using a batch size of \num{16384}, an initial learning rate of \num{5e-5}, and a \texttt{CosineAnnealingLR} learning rate scheduler.
    Additionally, we apply gradient clipping on the $\ell_2$ norm of the gradients.
    We only save the model that achieves the lowest validation loss.
    For FMPE, we train with automatic mixed precision (AMP) as it speeds up training significantly, while we found no such advantage for NPE.
    One training epoch takes \qty[per-mode=symbol]{328+-35}{\second\per\epoch} (total: 55~hours) for the FMPE model, while the average for the NPE model is \qty[per-mode=symbol]{975+-313}{\second\per\epoch} (total: 163~hours).
    All our models are trained on a single NVIDIA~H100 GPU.
    
    For the time prior $p(t)$ in the first expectation value of \cref{eq:fmpe-loss}, we found empirically that sampling $\sqrt[4]{t} \sim \mathcal{U}(0, 1)$ works well in practice.
    Compared to sampling $t \sim \mathcal{U}(0, 1)$, this approach places more emphasis on learning the vector field at later times, where $v_{t, x}$ is typically more complicated than at early times:
    Intuitively, at $t \approx 0$, each sample only needs to be moved approximately in the right direction, whereas at $t \approx 1$, each sample needs to be moved to its exact final location under the target distribution.
    Finally, we set $\sigma_\text{min} = \num{1e-4}$ for the $\sigma_\text{min}$ in \cref{eq:target-vectorfield}.

    \subsection{Inference procedure (for FMPE)}
    \label{app:subsec:inference-procedure}
    
    Using the FMPE model at inference time requires solving a differential equation; see \cref{eq:neural-ode,eq:cnf-density}.
    For this, we use the \texttt{dopri5} integrator as implemented by \texttt{torchdiffeq} \citep{Chen_2018b}.
    One critical parameter here is the tolerance: We found that a (relative and absolute) tolerance of \num{2e-4} is sufficient for sampling, while a value of \num{5e-5} is required for evaluating log-probabilities.
    Lower tolerances increase the runtime of the ODE solver, but do not improve the results (e.g., in terms of IS efficiency).

    \subsection{Test set generation}
    \label{app:subsec:test-set-generation}
    
    To generate a test set for our large-scale analysis, we randomly sample 500 values of $\theta$ from the Bayesian prior $p(\theta)$ and simulate the corresponding spectra, to which we add a random noise realization like in \cref{sec:simulator-setup}.
    We call this our \enquote{default} test set.
    When manually inspecting the spectra in the default test set, we noticed that a relatively large fraction (over \qty{20}{\percent}) have a mean noise-free flux value that is below the smallest assumed noise level.
    This means that, after adding noise, many of these spectra are virtually indistinguishable from pure noise, thus yielding the same posteriors.
    Similarly, there are also a considerable number of spectra with an average planet flux so high that they are effectively noise-free given the considered noise levels.
    We concluded that this might not be ideal for evaluation purposes and therefore decided to create another test set, consisting of spectra from a \enquote{neighborhood} around the benchmark spectrum: 
    We first map $\theta_0$ into the unit hypercube $[0, 1]^{16}$, using the inverse cumulative density functions of the priors.
    Based on the result $u_0$, we generate values of $\theta$ by sampling $z \sim \mathcal{N}(0, 0.1^2 \cdot I_{16})$, clipping $u = u_0 + z$ to the unit hypercube, and mapping $u$ back to the original parameter space.
    After running the simulator, we add a random noise realization in the same way as before.
    This results in spectra that are similar to the benchmark spectrum in terms of the mean and the total variation of the flux.
    Given that the $z$-vectors are drawn from a Gaussian, we call the result the \enquote{Gaussian} test set.

    \section{Nested sampling baselines}
    \label{app:nested-sampling}
    
    We use two Python-based implementations of nested sampling to create the baselines for our experiments in \cref{subsec:experiments-benchmark-spectrum}:
    
    \begin{description}[labelsep=0pt, itemsep=2mm]
        \item[\bfseries\nautilus] \citep{Lange_2023} is a recent implementation of nested sampling that utilizes importance sampling and makes use of neural networks to explore the target space more efficiently and improve the overall runtime.
        In our experiments, we reliably found \nautilus to be significantly faster than its competitors.
        We set the number of live points to \num{4000} and otherwise use the default settings of version 1.0.3 of the package.
        In particular, we leave the target number of posterior samples at \num{10000}, and note that increasing this value often results in super-linear increases in runtime.
        This is because the ESS does not grow monotonically with the number of proposals: drawing a posterior sample that increases the variance of the importance weights can also reduce the ESS.
        \item[\bfseries\MultiNest] \citep{Feroz_2009,Feroz_2019} is perhaps the most popular nested sampling implementation in the atmospheric retrieval community (see, e.g., \citealt{Alei_2024} or \citealt{Nasedkin_2024b} for recent examples) and was also used as a baseline by \citet{Vasist_2023}. 
        It is typically run using the \texttt{PyMultiNest} wrapper \citep{Buchner_2014}.
        We set the number of live points to \num{4000} and otherwise use the default settings of version 2.12 of the package.
        \texttt{PyMultiNest} does not provide an option to control the target number of posterior samples.
    \end{description}

    \noindent
    All nested sampling experiments were parallelized over 96 CPU cores.
    We also experimented with two more nested sampling implementations, which we did not end up including as baselines:
    
    \begin{description}[labelsep=0pt, itemsep=2mm]
        \item[\bfseries\dynesty] \citep{Speagle_2020} implements dynamic nested sampling to focus more on estimating the posterior instead of the Bayesian evidence \citep[see][]{Higson_2018}.
        After initial experiments using default settings resulted in poor estimates of the posterior (assessed by comparison with the results from other methods), we increased the number of live points to \num{5000} and set \texttt{sample="rwalk"} and \texttt{walks=100}.
        Despite a runtime of over 9 weeks, the resulting posterior was again in strong disagreement with the posteriors obtained by all other methods, leading us not to include the \dynesty results as a baseline.
        \item[\bfseries\UltraNest] \citep{Buchner_2021,Buchner_2022} implements a number of new techniques to improve the correctness and robustness of the results.
        However, after running \UltraNest with default settings for more than four weeks on our benchmark retrieval, the logs indicated that the method was still far from convergence and thus not feasible as a baseline.
    \end{description}

    \noindent
    For easier comparison, the (partial) results of all four nested sampling implementations on the noise-free benchmark spectrum from \cref{subsec:experiments-benchmark-spectrum} are summarized in \cref{tab:nested-sampling-stats}.
    We note further that \dynesty and \UltraNest are in good agreement with \nautilus and \MultiNest for test retrievals using only four atmospheric parameters.
    This leads us to believe that the issues we encountered in the full 16-dimensional case are likely due to a suboptimal choice of hyperparameters, and that a better configuration might yield better results.
    However, due to the high computational cost and long runtimes, optimizing the sampler settings is a substantial challenge in itself that is beyond the scope of this work.
    
    \begin{table*}[t]
        \centering
        \caption{
            Comparison of the four different nested sampling implementations on the noise-free benchmark retrieval from \cref{subsec:experiments-benchmark-spectrum}.
        }
        \label{tab:nested-sampling-stats}
        \begin{tabularx}{\linewidth}{
            X  %
            S[table-format=4.0]  %
            S[table-format={>\!\!}9.0]  %
            c  %
            S[table-format=6.0] %
            S[table-format=5.1] %
            S[table-format=2.1] %
            S[table-format={>\!\!}4.1]  %
        }
        \toprule
            \multicolumn{1}{l}{Method} & 
            \multicolumn{1}{c}{Live points} & 
            \multicolumn{1}{c}{Likelihood calls} & 
            \multicolumn{1}{c}{$\log Z$} & 
            \multicolumn{1}{c}{$N_\text{total}$} & 
            \multicolumn{1}{c}{$N_\text{eff}$} & 
            \multicolumn{1}{c}{$\varepsilon$ (in \unit{\percent})} &
            \multicolumn{1}{r}{Wall time (in \unit{\hour})} \\
        \midrule
            \dynesty\!\footnotesize{\textsuperscript{(1)}} & 
                5000 &  
                18730100 & 
                $401.497 \pm 0.080$ &
                206506 &
                41755.9 &
                20.2 &
                1548.1 \\
            \MultiNest & 
                4000 &
                17786920 &
                $407.182 \pm 0.080$ &
                30540 &
                \multicolumn{1}{c}{---} &
                \multicolumn{1}{c}{---} &
                326.2 \\
            \nautilus & 
                4000 &  
                1042944 & 
                $401.844 \pm 0.010$ &
                331008 &
                10004.4 &
                3.1 &
                19.5 \\
            \UltraNest\!\footnotesize{\textsuperscript{(2)}} &
                400 &
                {>\!\!}34319279 &
                \multicolumn{1}{c}{>\,213.0} &
                \multicolumn{1}{c}{---} &
                \multicolumn{1}{c}{---} &
                \multicolumn{1}{c}{---} &
                {>\!\!}813.5 \\
        \bottomrule
        \end{tabularx}
        \begin{FlushLeft}
            \textbf{Notes:}
            (1)~While the $\log Z$ estimate from \dynesty is similar to the other methods (including FMPE-IS / NPE-IS), the 1D and 2D marginal posteriors deviate strongly from all other methods.\quad
            (2)~We aborted the \UltraNest run when the number of likelihood evaluations exceeded the size of our ML training set and the sampler was still far from convergence (based on its $\log Z$ estimate).
        \end{FlushLeft}
    \end{table*}

    \section{Additional results}
    
\begin{figure*}[t]
        \centering
        \includegraphics[]{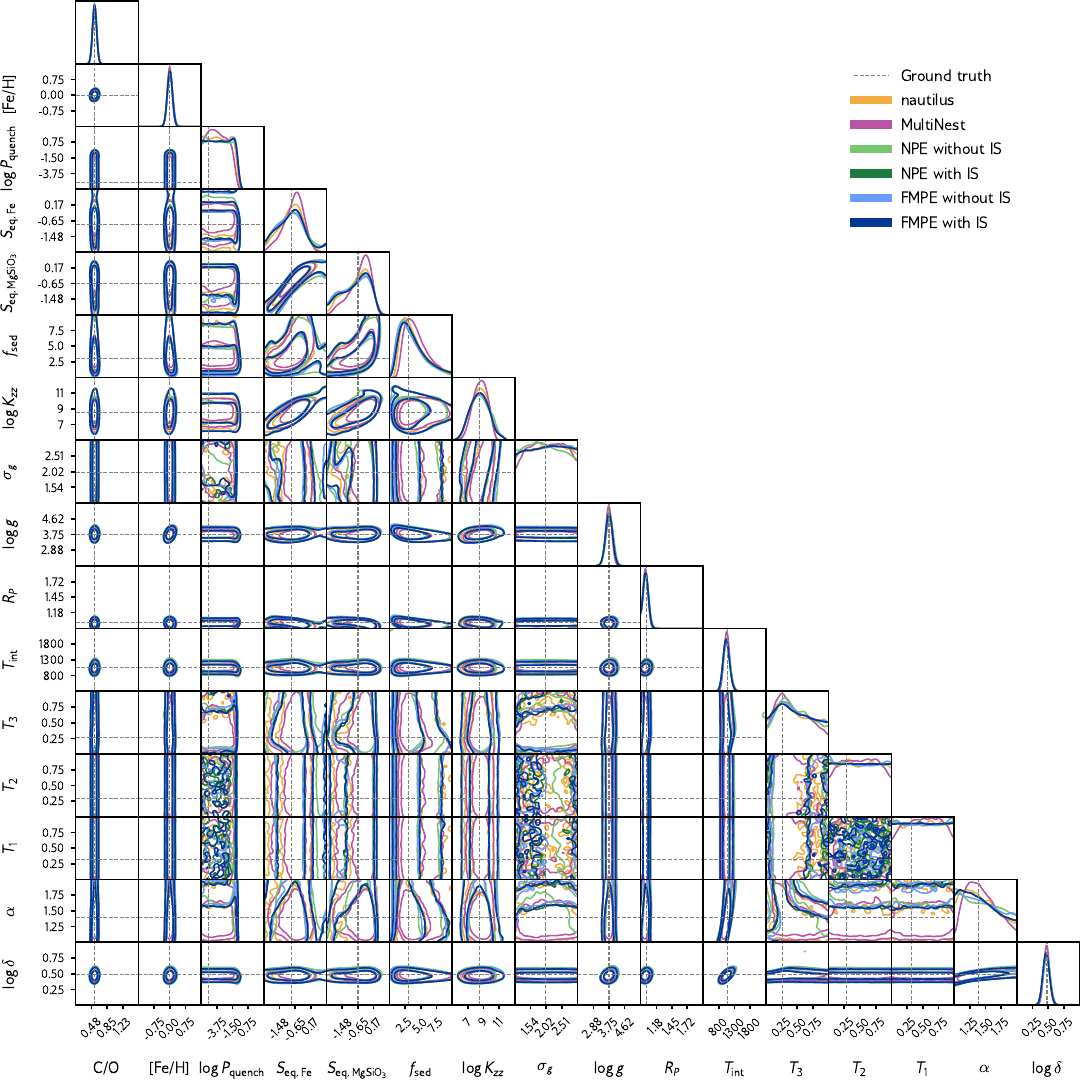}
        \caption{
            Cornerplot comparing the posteriors for the noise-free benchmark spectrum for all methods and all 16 parameters (cf. \cref{fig:cornerplot-subset}).
        }
        \label{fig:cornerplot-full}

    \end{figure*}

    \Cref{fig:cornerplot-full} contains the full 16-parameter version of \cref{fig:cornerplot-subset}.
    
    \section{Effect of training set size}
    \label{app:effect-of-training-set-size}

\begin{figure*}[t]
        \centering
        \begin{subcaptionblock}{0.5\linewidth}
            \centering
            \includegraphics[width=8cm]{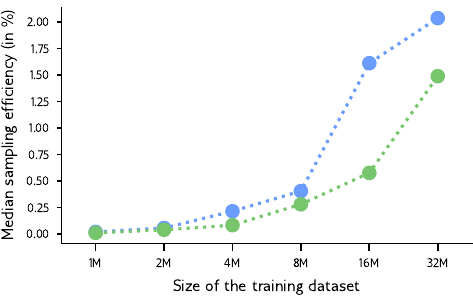}%
            \subcaption{Results on default test set.}
        \end{subcaptionblock}%
        \hfill%
        \begin{subcaptionblock}{0.5\linewidth}
            \centering
            \includegraphics[width=8cm]{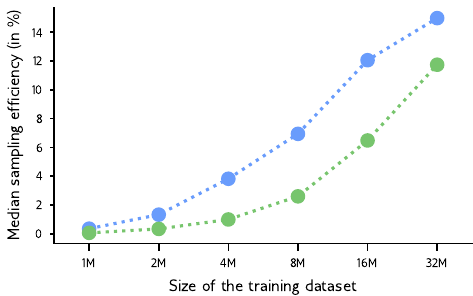}%
            \subcaption{Results on Gaussian test set.}
        \end{subcaptionblock}%
        \caption{
            \definecolor{C0}{HTML}{699CFC}
            \definecolor{C1}{HTML}{77C56D}
            Sampling efficiency as a function of the number of spectra used for training, for both
            \mbox{\raisebox{0.1ex}{\protect\tikz{\protect\draw[line width=1.5pt, draw=C0] (0, 0) -- (0.6, 0);\protect\draw[fill=C0, draw=none](0.3, 0) circle (0.11);}}}\,\textcolor{C0}{FMPE} and \mbox{\raisebox{0.1ex}{\protect\tikz{\protect\draw[line width=1.5pt, draw=C1] (0, 0) -- (0.6, 0);\protect\draw[fill=C1, draw=none](0.3, 0) circle (0.11);}}}\,\textcolor{C1}{NPE}.
            \vspace{1cm}
        }
        \label{fig:ablation-study}

        \label{LastPage}
    \end{figure*}

    We obtained the main results of this paper using ML models that were trained on a dataset of $2^{25} = \num{33 554 432}$ (``32M'') simulated spectra.
    To understand how the performance of FMPE and NPE depends on and scales with the amount of training data, we perform an ablation study where we retrain our models using subsets of the original training set consisting of $2^{24} = \num{16777216}$ (``16M''), $2^{23} = \num{8388608}$ (``8M''), $2^{22} = \num{4194304}$ (``4M''), $2^{21} = \num{2097152}$ (``2M'') and $2^{20} = \num{1048576}$ (``1M'') spectra, respectively.
    We train each model using the same settings as in the main part (see \cref{app:model-training} for details).
    For each model, we then run 100 retrievals on our two test sets (default and Gaussian).
    In total, this corresponds to $2\ \text{model types} \times 2\ \text{test sets} \times 5\ \text{new training set sizes} \times 100\ \text{test spectra} = \num{2000}$ additional retrievals.
    For each retrieval, we used \num{100 000} proposal samples for importance sampling and computed the sampling efficiency $\varepsilon$ as a metric for the quality of the proposal distribution.
    The results of this ablation study are shown in \cref{fig:ablation-study}.
    As expected, we observe a clear trend for both methods where the sampling efficiency increases with the size of the training set, which implies that the accuracy of the proposal distribution also increases with the amount of training data.

\end{document}